%% file: VO2_paper.tex
\documentclass[10pt]{iopart}
\usepackage{siunitx}
\usepackage[numbers]{natbib}
\usepackage{graphicx}
\usepackage[caption=false]{subfig}
\usepackage{multirow}
\usepackage{tablefootnote}
\usepackage{ulem}

\newcommand\tsub[1]{\textsubscript{#1}}
\newcommand\tsup[1]{\textsuperscript{#1}}
\DeclareGraphicsExtensions{.pdf,.jpeg,.png,.eps,.pdf}

\usepackage[dvipsnames]{xcolor}
 
\begin{document}

\title[Properties of the rutile VO\tsub{2}(110)  surface and its oxygen-rich terminations]{First principles studies of the electronic and structural properties of the rutile VO\tsub{2}(110)  surface and its oxygen-rich terminations}

\author{J Planer, F Mittendorfer, J Redinger}
\address{Center for Computational Materials Science, Institute of Applied Physics, Vienna University of Technology, Vienna}
\ead{florian.mittendorfer@tuwien.ac.at}

\begin{abstract}
We present a Density Functional Theory (DFT) study of the structural and electronic properties of bare rutile VO\tsub{2} (110) surfaces and its oxygen-rich terminations.
We discuss the performance of various DFT functionals, including PBE, PBE+U (U = \SI{2}{eV}), SCAN and SCAN+rVV functionals with non-magnetic and ferromagnetic spin ordering. 
We predict the presence of a ring-like termination that is electronically and structurally related to a V\tsub{2}O\tsub{5} (001) monolayer and shows a higher stability than pure oxygen adsorption phases. 
Despite the fact that the calculated phase stabilities depend on the chosen functional, our results show that employing the spin-polarized SCAN functional offers a good compromise yielding both a reasonable description of the structural and electronic properties of  the rutile VO\tsub{2} bulk phase and the enthalpy of formation for different stages of vanadium oxidation.  
\end{abstract}
\submitto{\JPCM}
\maketitle

%
%
%
%
\ioptwocol
\input{introduction}
\input{comp_details}

\input{results}

\input{summary}
\input{acknowledgement}

\input{bibliography}
\end{document}


\title[Properties of the rutile VO\tsub{2}(110)  surface and its oxygen-rich terminations]{First principles studies of the electronic and structural properties of the rutile VO\tsub{2}(110)  surface and its oxygen-rich terminations}

\author{J Planer, F Mittendorfer, J Redinger}
\address{Center for Computational Materials Science, Institute of Applied Physics, Vienna University of Technology, Vienna}
\ead{florian.mittendorfer@tuwien.ac.at}
\submitto{\JPCM}
\maketitle

\section{Supplementary information}
\subsection{Generation of the ($2\times 2$) superstructures}
In order to sample the large configuration space of the VO\tsub{2} (110) ($2\times 2$) terminations, we performed an optimization of the random structures generated by the USPEX package\cite{Oganov2006,Lyakhov2013,Oganov2011}, which comprised of the following steps: \begin{itemize}
	\item 
	Generation of the random structures at a given surface stoichiometry with the USPEX package\cite{Oganov2006,Lyakhov2013,Oganov2011}, which is based on an evolutionary algorithm developed by Oganov, Glass, Lyakhov and Zhu and features a local optimization, a real-space representation and flexible physically motivated variation operators. We generated 1000 initial structures separately for the PBE, PBE+U and the SCAN functional. The samples consisted of the surface layer and one bottom layer.
	\item
	The initial structures were pre-optimized within 170 ionic steps with the electronic loop converged to the \SI{e-3}{eV}, performing non-spin polarized calculations. The energy cut-off was set to \SI{400}{eV} and the k-point grid was downgraded to the $\Gamma$-point.
	\item
	We chose the 100 most stable structures from the previous step, which were then relaxed using more accurate settings: an \SI{450}{eV} energy cut-off, a $2\times1\times1$ k-points grid, an electronic loop converged to \SI{e-5}{eV}. The ionic relaxation was stopped when all residual forces were smaller than \SI{e-2}{meV\per\AA}.
	\item
	The 30 most stable structures were selected from the previous step and augmented by an additional substrate layer. In these calculations, only the bottom layer was fixed to its bulk positions, the surface and sub-surface layer were allowed to relax.
	\item
	Up to this point, we performed the optimization steps separately for the PBE, PBE+U and the SCAN functionals. Since the  resulting structures for the different functionals deviated from each other, we chose the 10 most stable structures from the pool obtained when applying all the functionals. These structures were subsequently further optimized with the respective functionals.
	\item
	We selected approximately the 10 most-stable structures for all functionals and increased their slab thickness by adding two more layers where the bottom layer was again fixed at the bulk positions. The lateral lattice vectors were adjusted to the bulk values obtained by the respective functional and the k-points grid was increased to $4\times2\times1$. At this point we also considered spin-polarized and SCAN+rVV10 calculations for these ten most stable structures. 
	\item
	In the final step, we switched to symmetric five-layered slabs to overcome potential errors due to the occurrence of dipoles in  asymmetric slabs. The same computational settings as in the previous step were  used.   
\end{itemize}

\begin{figure*}
\centering
\subfloat [][]{\label{fig:uspexProcess_step_1}%
  	\includegraphics[width=.24\textwidth]{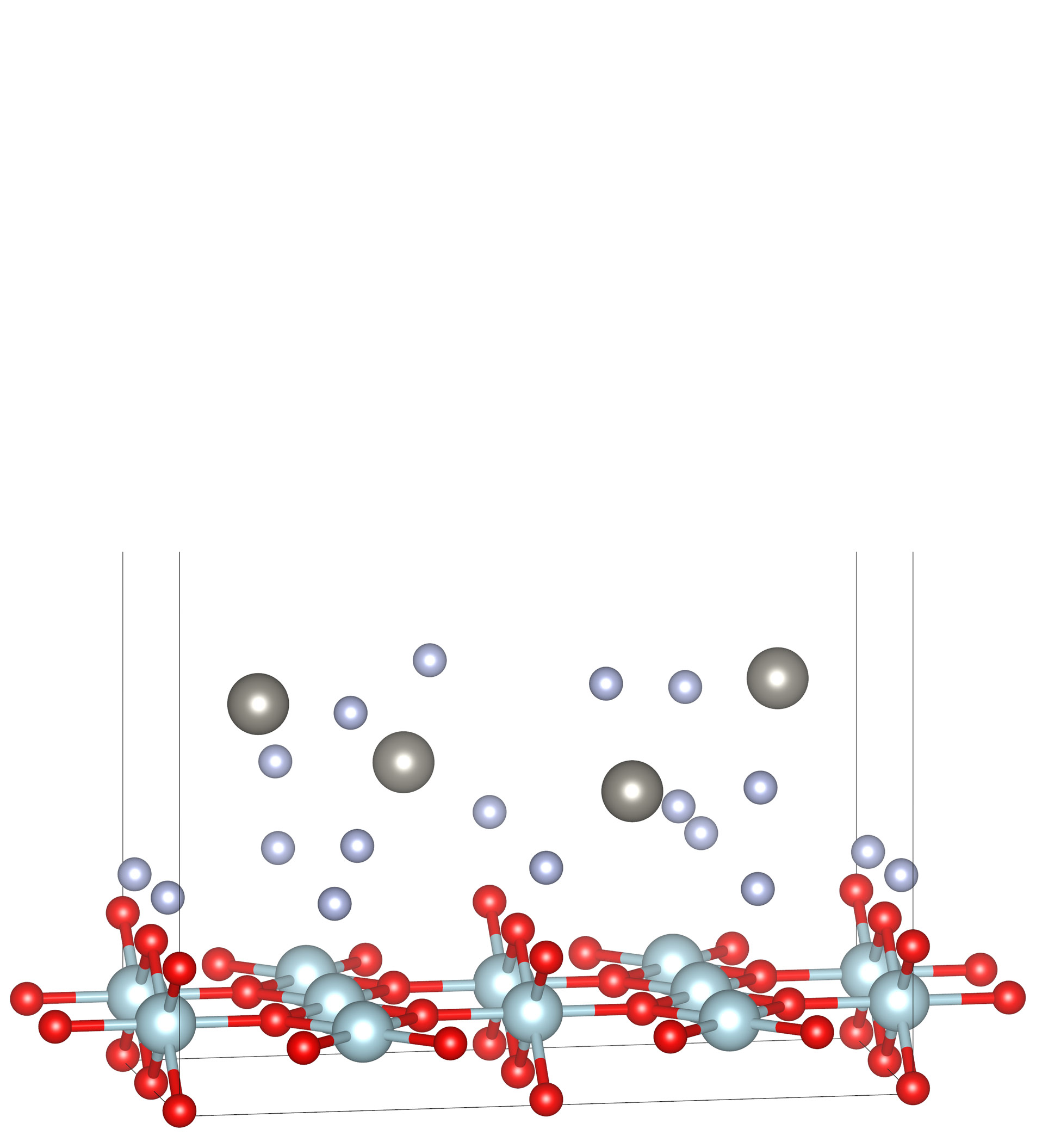}  
	}
\subfloat[][]{\label{fig:uspexProcess_step_2}%
 	\includegraphics[width=.24\textwidth]{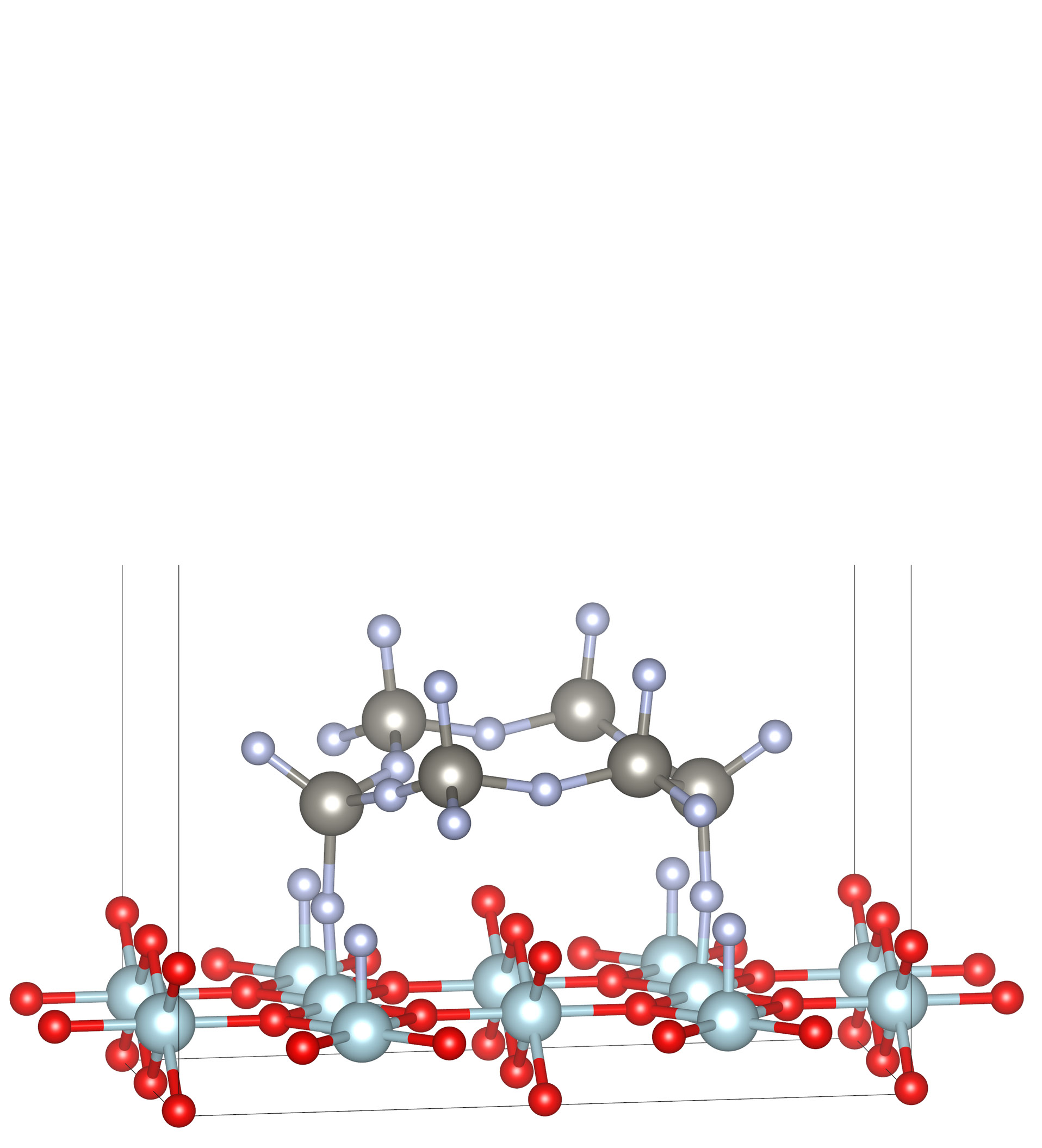}  
         }
\subfloat[][]{\label{fig:uspexProcess_step_3}%
  	\includegraphics[width=.24\textwidth]{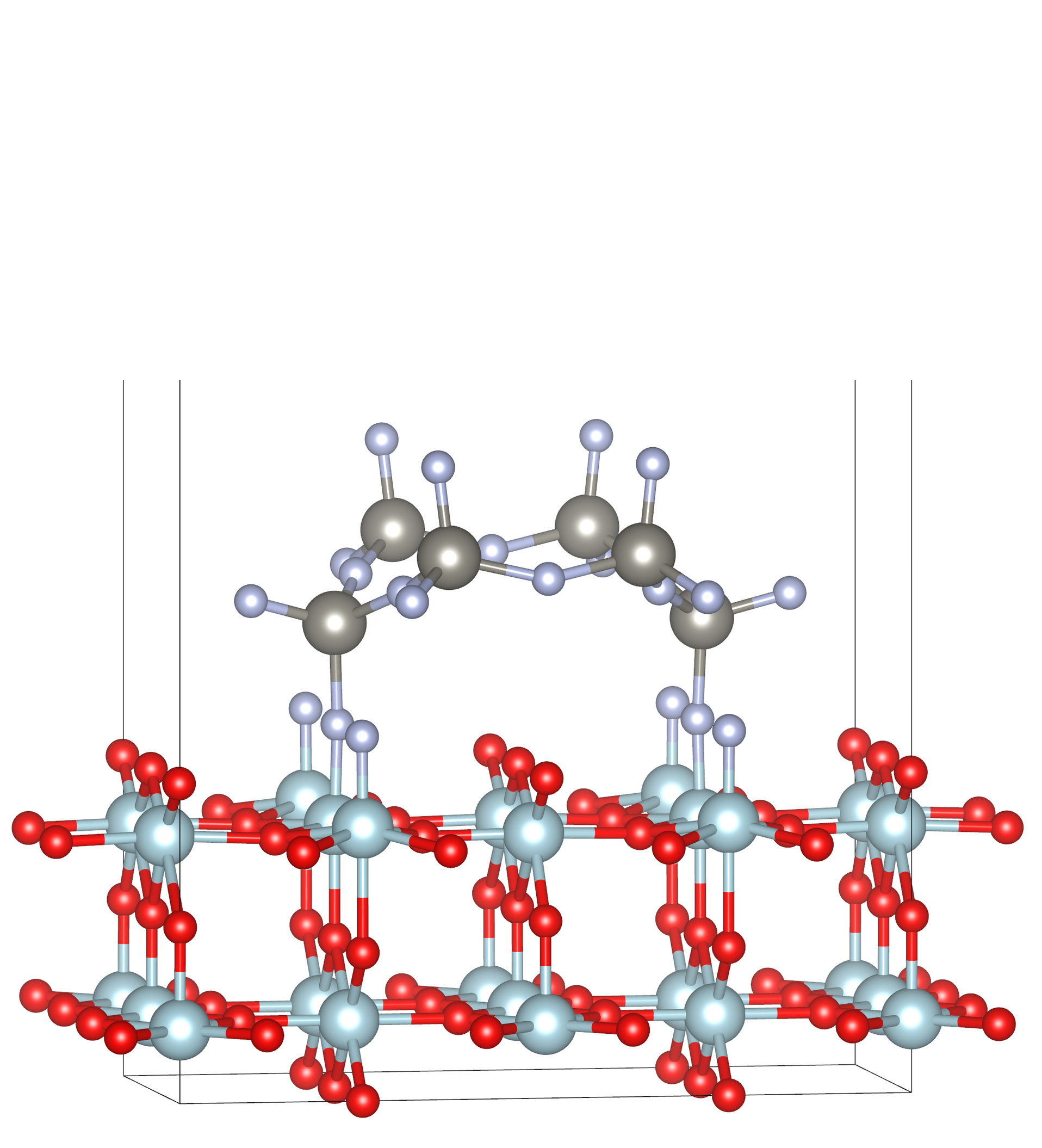}  
  	}
\subfloat[[][]{\label{fig:uspexProcess_step_4}%
  	\includegraphics[width=.24\textwidth]{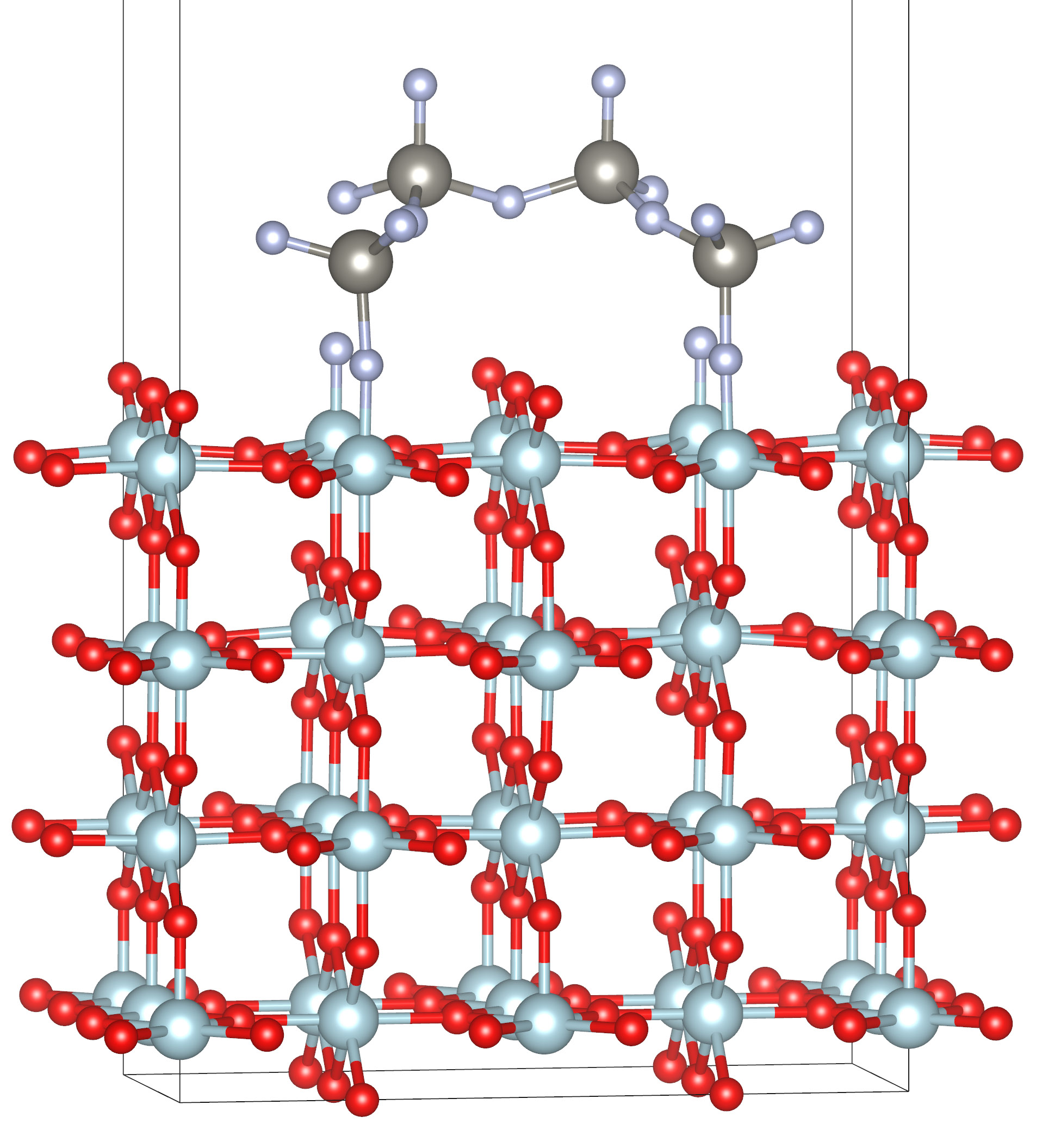} 
  	}
\caption{Optimizaton process of the random structures. At the beginning, a set of 1000 random structures was generated with the USPEX\cite{Oganov2006} package as it is shown in panel (a). After that, all of them were pre-optimized, and the \SI{10}{\percent} most stable ones fully relaxed, which led to a set of 100 structures, similar to the one from panel (b). At the next step, an additional substrate layer was added as shown in panel (c). Since two substrate layers were too thin according to our convergence studies, the last step consisted of adding two more substrate layers, as shown in panel (d). Concluding, a full and accurate relaxation with all functionals using both spin-polarized (FM) and non-spin polarized (NM) electronic configurations was performed.}
\label{fig:uspexProcess}
\end{figure*}      

\par
The optimization process of the random structures is depicted in Figure \ref{fig:uspexProcess}. Violet and gray spheres represent the oxygen and vanadium atoms that form the surface layer respectively. In our notation, V\tsub{8}O\tsub{16} is the stoichiometry of the single VO\tsub{2} layer. First we  considered V\tsub{8}O\tsub{18} and V\tsub{8}O\tsub{20} surface stoichiometries, which ist the the same as for the whole VO\tsub{2} trilayer with 2 and 4 additional oxygen atoms, respectively. With our approach we obtained the fully and half oxygen covered (110) surface respectively and therefore concluded that the configuration space is sampled sufficiently well, even though there is no guarantee for finding at the given stoichiometry the most stable structure. In our work we first considered  strongly oxidized surface stoichiometries V\tsub{$n$}O\tsub{2$n$+2} to V\tsub{$n$}O\tsub{2$n$+5} for $n$ ranged from 4 to 8 which corresponds to the more than a half of an additional VO\tsub{2} trilayer. Nevertheless, in the case of surface terminations with $n \geq 6$, we either observed formation of the second surface layer or the formation of a (110) termination with a V vacancy. Both these superstructures turned out to be unstable with respect to the bare (110) surface and other ring-like terminations.    

\subsection{Simulated STM images}
\begin{figure}[!h] 
\centering
\includegraphics[width=\linewidth]{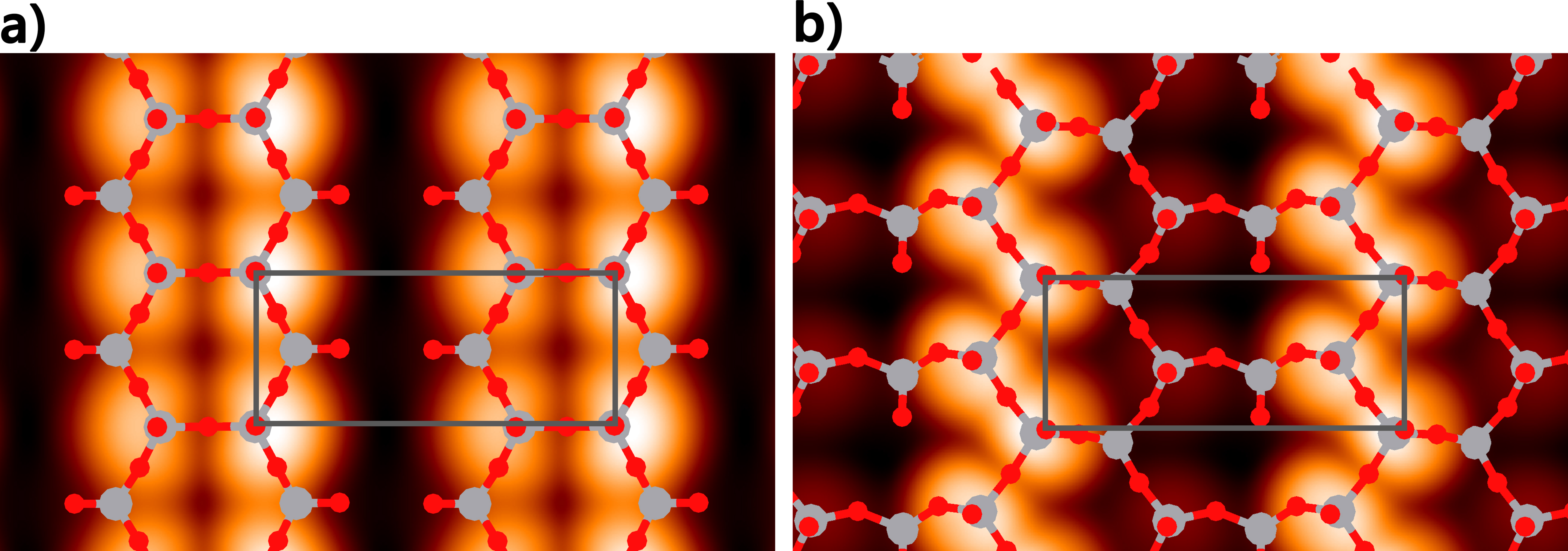}
\caption{Simulated STM images (empty states) of the ring terminations with the (a) V\tsub{4}O\tsub{13} [Fig.~8(b,c)] and (b) V\tsub{5}O\tsub{15} [Fig.~8(d)]  surface stoichiometries.}
\label{fig:DOSR}
\end{figure}

\clearpage
\subsection{Wannier-projected density of states}
\begin{figure}[!h] 
\centering
\includegraphics[width=0.9\linewidth]{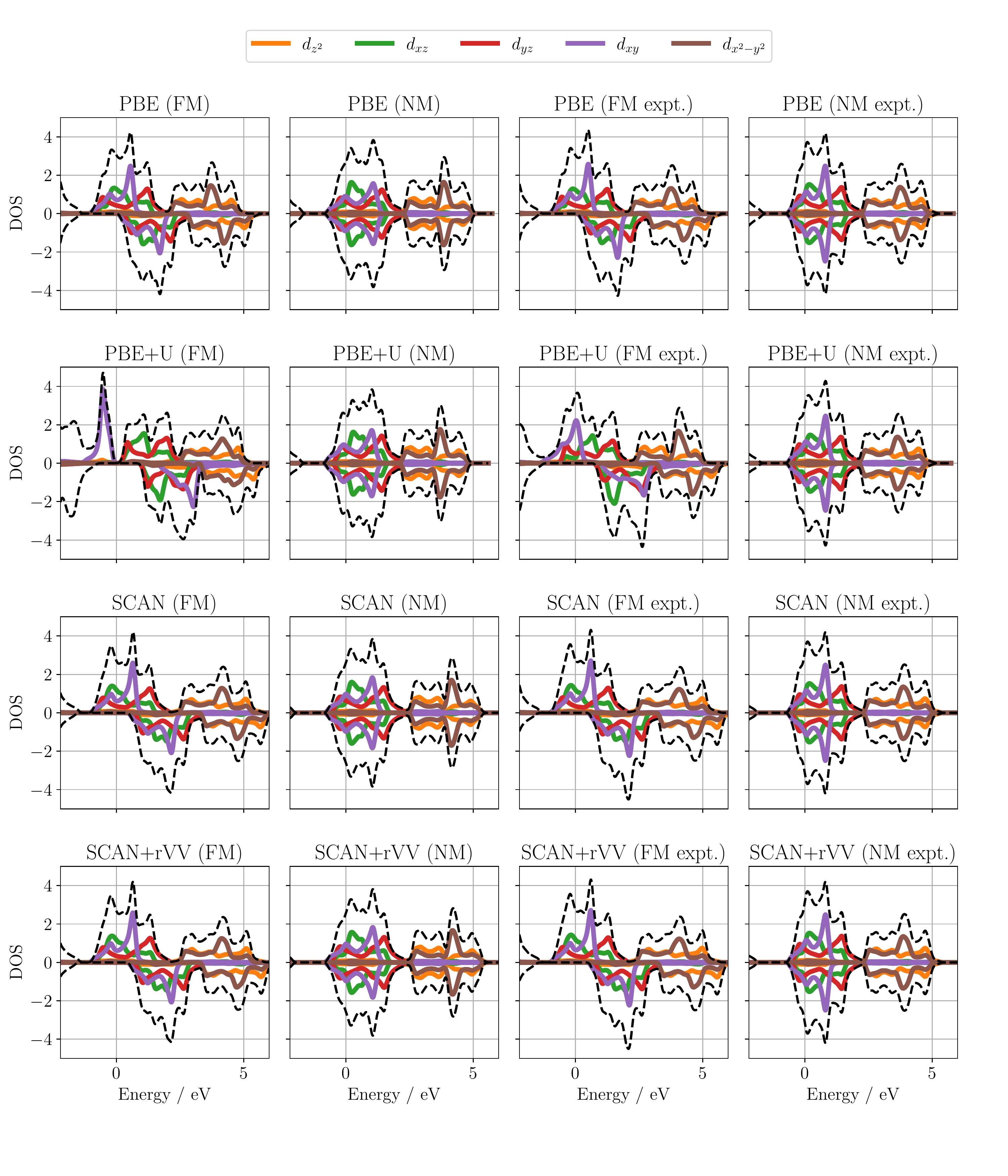}
\caption{Projected density of states onto Wannier orbitals localized at vanadium atoms in the rutile (R) phase, calculated with several DFT functionals using relaxed (two left columns) and experimental structures (two right columns). 0 marks the Fermi energy, $t_{2g}$ states are formed by $d_{xz}$, $d_{yz}$ and $d_{xy}$ orbitals, $d_{z^2}$ and $d_{x^2-y^2}$ form the $e_g$ states. }
\label{fig:DOSR}
\end{figure}
\clearpage
\input{bibliography}

%% file: introduction.tex
\section{Introduction}
Vanadium dioxide undergoes a metal-to-insulator transition (MIT) at \SI{68}{\celsius} which makes it usable in many applications such as optoelectronic switches, sensors or Mott-field effect transistors\cite{Stefanovich2000,Strelcov2009,Sengupta2011}. Below the transition point, $\mathrm{VO_2}$ undergoes the phase transition from a metallic rutile (R) phase (space group P4\tsub{2}/mnm) to a semiconducting monoclinic (M) phase (space group P2\tsub{1}/c), forming chains of paired vanadium atoms and thereby diminishing its electrical conductivity. The rutile phase is paramagnetic above the transition temperature while the monoclinic phase is considered to be nonmagnetic\cite{Hiroi2013}. The transition temperature can be driven towards  
room temperature by doping with high valence cations such as $\mathrm{W^{(6+)}}$ \cite{Netsianda2008} to make this material available in real applications, such as smart window coatings.
\par
Owing to the numerous potential applications of VO\tsub{2}, the bulk properties have been studied applying a vast number of theoretical approaches\cite{Zhu2012}. It has been shown that standard methods within the DFT framework fail to completely
recover the electronic, structural and magnetic properties of the rutile and monoclinic VO\tsub{2} phase. For example, both the  local density approximation (LDA) and the generalized gradient approximation (GGA) fail to open an \SI{0.6}{eV} band gap in the monoclinic (M1) phase\cite{Eyert2002,Wentzcovitch1994}. This shortcoming might be eliminated either by using the DFT+U variant \cite{Liu2010}, or meta-GGA and hybrid functionals, as analyzed by Stahl and Bredow \cite{Stahl2019,Bredow2018}. According to their findings, DFT+U leads to strong structural distortions of the monoclinic phase and a wrong energetic ordering of the VO\tsub{2} phases. 
Furthermore, conventional hybrid functionals such as HSE06 or PBE0 lead to a splitting 
of the conduction band in the metallic rutile phase and thus the Fock mixing parameter needs to be adjusted to correctly describe the electronic structure of the rutile and monoclinic phases\cite{Pantelides2017,Bredow2018}. Stahl and Bredow  also concluded\cite{Stahl2019} that the meta-GGA SCAN functional offers a good compromise between accuracy and  computational cost. 
\par
The driving mechanism behind the MIT was recently investigated by Brito et al.\cite{Brito2016} by performing combined DFT and embedded dynamical mean-field theory (DMFT) calculations. 
Such a  combination constitutes a powerful approach to evaluate the many-body electronic structure of strongly correlated materials\cite{DMFTReview2006} and in particular to describe the MIT between a metal and a Mott insulator, as is the case for the present system, where the electronic transition can be described as a Mott transition in the presence of strong intersite exchange.
This is in agreement with the finding of Zhu et al.\cite{Zhu2012} who used the modified Becke-Johnson exchange together with LDA correlation potentials for both 
rutile and monoclinic VO\tsub{2} phases, to show that the MIT can be characterized as a correlation driven transition.
\par
In contrast to  VO\tsub{2}  bulk systems, 
VO\tsub{2} surfaces and VO\tsub{2} thin films have been studied to a much lesser degree.
The surface free energy of bare and oxygen-covered low-index facets was
investigated  by Mellan et
al.\cite{Mellan2012} employing the non-magnetic PBE functional. 
PBE+U calculations  performed by Wahila et
al.\cite{Wahila2020}, indicate that the surface free energy of corresponding surfaces is lower for the rutile 
than for the monoclinic phase.
Furthermore, they find  it necessary to include spin-polarization in order to avoid 
negative values for the surface free energies of oxygen-rich reconstructions. 
In a very recent study\cite{stahl_surfaces_2021}, the rutile and monoclinic VO\tsub{2} surfaces were studied with the hybrid sc-PBE0 functional with an adjusted Fock mixing parameter. Despite its good performance on 
both bulk phases, the hybrid sc-PBE0 functional surprisingly fails to describe the surface energies of rutile VO\tsub{2} which do not converge upon increasing the thickness of the slab and 
even yield negative values for thicker slabs. 
On the
experimental side, recent studies show the presence of a ($2\times 2$) reconstruction on the VO\tsub{2} (110) surface, accompanied by a change in the 
surface oxygen content  of single crystals\cite{Wagner20} and thin 
films \cite{Fischer2020,Wahila2020}. 
\par
The aforementioned DFT studies using the PBE(+U) approach on the rutile VO\tsub{2} (110)
surfaces under oxygen-rich conditions \cite{Wahila2020,Mellan2012} were 
done only for regular oxygen adsorption phases.   
However, a very recent study by Wagner et al. \cite{Wagner20} indicates the
presence of ($2\times2$) surface reconstructions with tetrahedrally coordinated V atoms. 
In order to shed some more light on these findings,
we present a detailed study of various oxygen adsorption phases 
and
surface reconstructions on rutile VO\tsub{2} surfaces, and evaluate the performance of common DFT
functionals, including GGA, GGA+U,meta-GGA and meta-GGA+U.
In agreement with the experimental findings of Wagner et al. \cite{Wagner20}, we
show that the simple regular oxygen adsorption phases considered in the previous
calculations\cite{Mellan2012,Wahila2020} are less stable than 
surface terminations with tetrahedral V coordination polyhedra that are structurally and electronically related to a
V\tsub{2}O\tsub{5} (001) monolayer.
This findings impose another important constraint to the DFT functional used for the description of off-stoichiometric
VO\tsub{2} surface terminations, namely a correct description of the stability of the insulating 
V\tsub{2}O\tsub{5} phase with respect to VO\tsub{2}.

%% file: comp_details.tex
\section{Computational details}
All calculations were performed with the \textit{Vienna Ab-initio Simulation Package} (VASP) \cite{Kresse1996,Zhang1996}. We used the projector augmented wave (PAW) method \cite{Blochl1994,Joubert1999} for treating the core electrons. The Bloch-functions for the 6 valence electrons of oxygen (2s\textsuperscript{2}2p\textsuperscript{4})  and the 13 for vanadium (3s\textsuperscript{2}3p\textsuperscript{6}3d\textsuperscript{4}4s\textsuperscript{1}), were expanded in a plane wave basis set with an energy cut-off \SI{500}{eV} for all final relaxations. The Brillouin zone was sampled with the $\Gamma$-centered Monkhorst-Pack scheme \cite{Pack1977} 
using \num{1,4} k-points\slash\AA$^{-1}$
to ensure the absolute convergence within the accuracy \SI{1}{meV} per atom. The electronic optimization self-consistent loop was converged to the \SI{e-5}{eV} and ionic relaxation was stopped when residual forces were smaller 
than \SI{e-2}{\eV \per \angstrom}.
\par
The local electronic structure was also studied with help of the Wannier90 code \cite{Pizzi2020}. All sub-bands  that comprise O $2p$ and V $3d$ bands were projected onto the same amount of Wannier functions. The local coordinate system of the Wannier functions localized at vanadium atoms was rotated by \SI{45}{\degree} around the $z-$axis as proposed by Eyert \cite{Eyert2002} in all systems studied. This rotation principally also implies an exchange of $d_{xy}$ and $d_{x^2-y^2}$ orbitals, but in our work we keep the naming of the orbitals the same as in the original basis. 
\par
We used several DFT functionals to describe the exchange-correlation energy, namely the PBE,  PBE+U, SCAN\cite{Sun2015}, SCAN+U and SCAN+rVV \cite{Peng2016}. The influence of the Hubbard U within Dudarev's implementation \cite{Dudarev1998} on the properties of VO\textsubscript{2} phases has been studied in the work of Stahl et al \cite{Stahl2019}, showing that values of U between \SIrange{1.4}{2.6}{eV} correctly open a band gap in the monoclinic phase, but not in the rutile phase if  the experimental structures are used in the calculations. We therefore settle on an U value of \SI{2}{eV} as  used before in point defect calculations \cite{Wickramaratne2019}. Based on these functionals we performed both non-spin polarized (NM) and spin-polarized calculations of ferromagnetic arrangements (FM) which turned out to be energetically most favourable for the rutile phase \cite{Stahl2019}.
\par
All slab calculations were performed at bulk optimized lattice constants applying the corresponding DFT functional and spin ordering. The stability of the respective VO\tsub{2} terminations was determined via evaluation of the surface free energy, calculated by linear regression from total energies of 5 to 8-layered slabs, as  described by the following scheme. The surface energy of a slab that is composed of $N$ bulk formula units, having a surface area $S$ is extracted from DFT calculations according to the known formula:
\begin{equation}
	\sigma = \frac{E_\mathrm{slab}-N\cdot E_{\mathrm{bulk}}}{2\cdot S},
\end{equation} 
where $E_\mathrm{slab}$ and $E_{\mathrm{bulk}}$ are the total slab and bulk energies obtained from DFT calculations. A direct evaluation of $\sigma$ requires a different sampling of Brillouin zones for bulk and slab models. In order to avoid this drawback, this formula can be rearranged towards a slab energy as a function of the number of layers, 
\begin{equation}
	E_\mathrm{slab}=E_\mathrm{slab}(N) = k\cdot N + q.
\end{equation} 
Comparison of these equations yields an expression for the surface energy as:
\begin{equation}
	\sigma = \frac{q}{2S},
\end{equation}
where q is a parameter obtained from a linear regression based on calculations for several slab thicknesses.
\par
To calculate the surface free energy of the most stable off-stoichiometric terminations we follow the procedure described in the work of Reuter \cite{Reuter2001}. We approximated the Gibbs free energy with the total DFT energies of the slabs and reference systems in the oxygen-rich limit so that all vibrational contributions are neglected. The  formula used to extract the surface free energy from a slab that consists of $N_{\mathrm{V}}$ vanadium atoms and $N_{\mathrm{O}}$ oxygen atoms as a function of oxygen chemical potential $\mu_{\mathrm{O}}$ reads
\begin{equation}
\gamma = \frac{E_{\mathrm{slab}}-N_{\mathrm{V}}\cdot E_{\mathrm{VO_2}} - \left( N_{\mathrm{O}} - 2N_{\mathrm{V}}  \right)\cdot\left( \frac{1}{2}E_{\mathrm{O_2}}+\mu_{\mathrm{O}} \right)}{2S}, 
\end{equation}
where $E_{\mathrm{VO_2}}$ and $E_{\mathrm{O_2}}$ are reference energies for the total energy of the rutile VO\tsub{2} bulk and oxygen molecule respectively, $E_{\mathrm{slab}}$ marks the total energy of the slab with the surface area $S$. We used 5-layered slabs with the symmetric top and bottom surfaces with a separating vacuum layer kept at \SI{15}{\AA} to study all ($2\times2$) surface reconstructions. 
The values of the oxygen chemical potential are considered for  a range of energies where the VO\tsub{2} phase is thermodynamically stable. Its oxygen-rich limit is approximately given by the reaction enthalpy of the oxidation reaction of the VO\tsub{2} phase:
\begin{equation}
\mathrm{2VO_2 + 1/2O_2 \rightarrow V_2O_5}.
\end{equation}
 The experimental value \SI{-1.28}{eV} is determined from the formation enthalpies
 of the VO\tsub{2}\cite{LandoltBornstein2001} and
 V\tsub{2}O\tsub{5}\cite{LandoltBornstein2001_2} while the calculated value is
 considered in the \SI{0}{K} temperature limit by neglecting vibrational and
 entropic contributions. This approximation yields an \SI{1.3}{\percent} error in
 experimental heat of formation of vanadium pentoxide compared to the value
 obtained at the room temperature\cite{LandoltBornstein2001_2}. The reaction
 enthalpy
 is therefore calculated using the total energies of bulk systems and oxygen molecule as
 \begin{equation}
 H_{\text{f}}^{\text{DFT}}(\mathrm{V_2O_5}) = E^{\text{bulk}}_{\mathrm{V_2O_5}} - (2E^{\text{bulk}}_{\mathrm{VO_2}}+1/2E_{\mathrm{O_2}}).
\end{equation} 
\par
In order to sample the large configuration space of the VO\tsub{2}(110) ($2\times 2$) terminations, we performed an optimization of the random structures generated by the USPEX package \cite{Oganov2006,Lyakhov2013,Oganov2011}. A more detailed information on this optimization procedure can be found in the supplementary material.

%% file: results.tex
\section{Results}
\subsection{Bulk properties of the VO\tsub{2} (R) and V\tsub{2}O\tsub{5} phases}
\begin{figure}
\includegraphics[width=\linewidth]{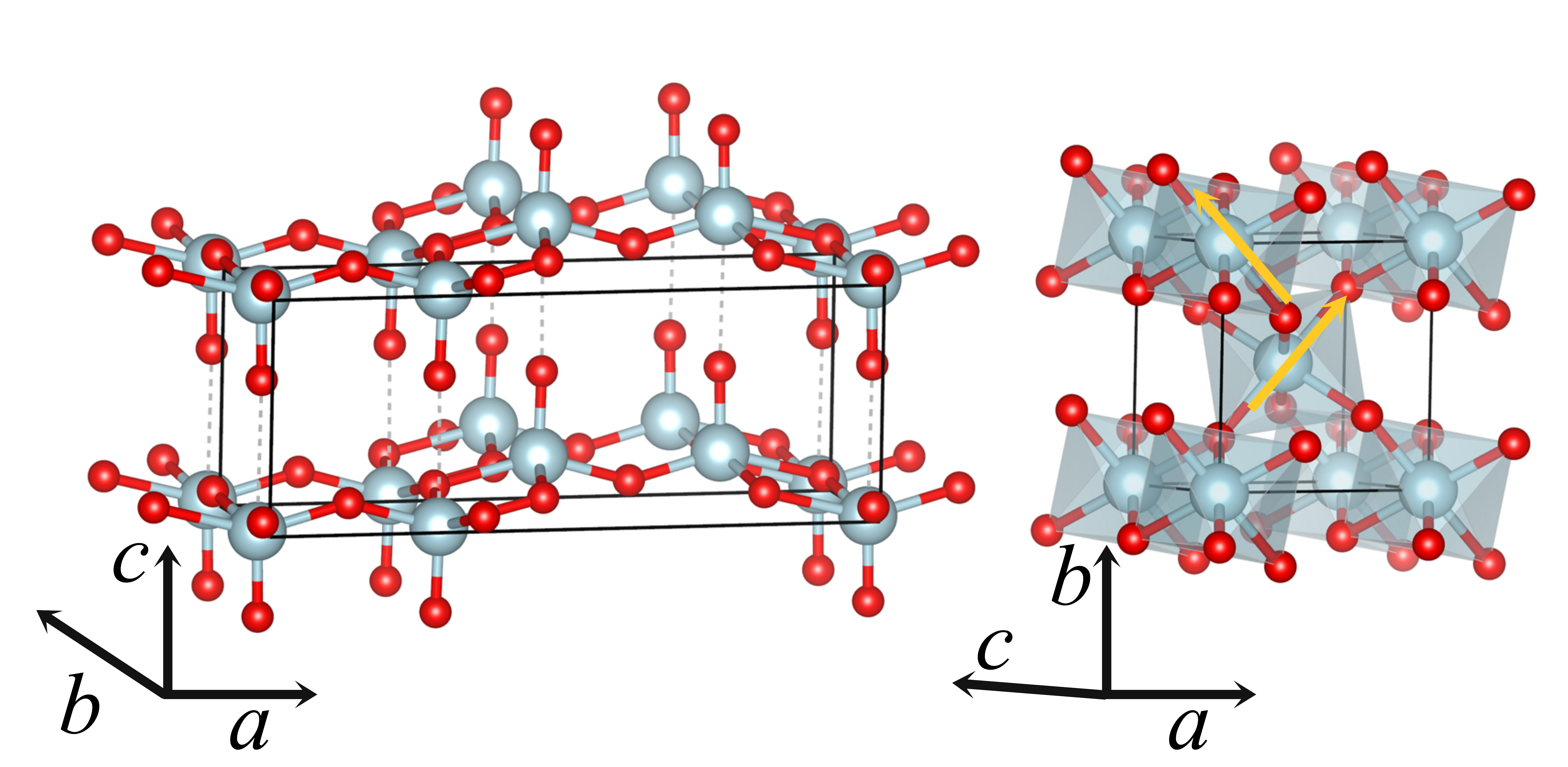}
\caption{The structures of V\tsub{2}O\tsub{5} (left panel) and rutile VO\tsub{2} (right panel). The dashed lines in the left panel mark the  inter-layer van der Waals bonds along the \textit{c}-direction between V\tsub{2}O\tsub{5} layers. The yellow arrows in the right panel denote the \textit{z}-directions  of the individual VO\tsub{6} octahedra's  local coordinate systems, which are mutually orthogonal to the  \textit{c}-direction of  the rutile VO\tsub{2} unit cell.}
\label{fig:strucs}
\end{figure}

The high-temperature rutile VO\tsub{2} phase exhibits a tetragonal structure
(space group P4\tsub{2}/mnm, a=\SI{4.55}{\AA},
c=\SI{2.86}{\AA})\cite{Rogers1993} with two VO\tsub{2} units that form the
unit cell. Oxygen atoms are arranged in octahedra around vanadium atoms with a slightly shortened z-axis that is arranged perpendicular to the
\textit{c}-lattice vector as visualized in Figure \ref{fig:strucs} (right panel). In contrast, the V\tsub{2}O\tsub{5} phase shows an 
orthorhombic structure (space
group P\tsub{mmn}, a=\SI{11,51}{\AA}, b=\SI{3,56}{\AA},
c=\SI{4,37}{\AA})\cite{Enjalbert1986} that is composed of van der Waals bonded layers stacked along
 the [001] direction. Note that there
is certain ambiguity in the notation of the lattice vectors and in some papers the 
\textit{b-} and \textit{c-}lattice vectors are interchanged. In the present work
we consider the V\tsub{2}O\tsub{5} layers to be oriented  parallel to the (001) plane as shown in the left panel of Figure \ref{fig:strucs}.
\par
While the structural and electronic properties of the rutile VO\tsub{2} phase were calculated using both ferromagnetic and non-magnetic spin configurations, only closed-shell calculations were performed for the V\tsub{2}O\tsub{5} phase because no unpaired electrons are present. Both phases were treated with all the aforementioned DFT functionals. The results are summarized in Table \ref{res_bulkLat_R}. Considering the rutile VO\tsub{2} phase, the lattice parameters are found to be well described well both with spin-polarized PBE and SCAN(+rVV) functionals, leading to errors below \SI{1}{\percent}, while PBE+U and SCAN+U functionals tend to overestimate the \textit{c}-lattice constant by \SI{6}{\percent}, causing a similar overestimation of the \textit{c/a} ratio. On the other side, non-magnetic 
calculations underestimate the \textit{c-}lattice vector approximately by $\sim$\SI{3}{\percent} while the \textit{a}-lattice vector turns out to be either overestimated by  \SI{2}{\percent} (PBE and PBE+U) or in good agreement with the experimental values (SCAN and SCAN+rVV). Consequently, the \text{c/a} ratio is calculated as 3-\SI{5}{\percent} lower than the experimental value.
\par
Concerning the V\tsub{2}O\tsub{5} phase, all functionals yield  in-plane \textit{a}- and \textit{b}-lattice parameters 
in very good agreement with experiment, with relative errors below \SI{1.5}{\percent}. For the \textit{c}-lattice vector perpendicular to the layers the agreement is worse, being either overestimated  by
\SI{11}{\percent} for PBE and PBE+U, or underestimated by
\SI{3}{\percent} for SCAN, by \SI{4}{\percent} for SCAN+U and \SI{5}{\percent} for SCAN+rVV. An overestimation by
\SIrange{6}{8}{\percent} in case
of  the PBE functional 
 has already been reported in previous
studies\cite{Goclon2009,Kresse2001,das_structural_2019}.
According to our calculations, a slightly enlarged interlayer distance is accompanied only
by a small energy penalty and thus our values  differ slightly from the
reported values while the \textit{a}- and \textit{b}-lattice constants match very well.
It should also be noted that the spacing between the V\tsub{2}O\tsub{5} layers 
is already underestimated on the SCAN
level, and that the additional van der Waals contributions in the SCAN+rVV
functional lead to an even smaller value. 
Concerning the unit cell volumes of rutile VO\tsub{2}, non spin-polarized calculations as expected deliver
slightly  smaller volumes than spin-polarized calculations. 
Comparing the cell volumes of  rutile VO\tsub{2} and V\tsub{2}O\tsub{5}, the additional 
oxygen and the concomitant transformation to the layered structure of the V\tsub{2}O\tsub{5} phase leads to an increase of the volume per formula unit by a factor of \SI{1,5} to \SI{1,7}. Similar to the \textit{c}-lattice vector,  the calculated volume is either underestimated by  \SI{2}{\percent} for SCAN and  SCAN+U, and by \SI{4}{\percent} for SCAN+rVV or overestimated by \SI{12}{\percent} for PBE and \SI{13}{\percent} for PBE+U.
\par
One of the fundamental  properties of the electronic structure is the value of the band gap between the valence and the conduction band. The rutile VO\tsub{2} phase is known to be metallic, while V\tsub{2}O\tsub{5} is a semiconductor with an electronic band gap of about \SI{2.3}{eV}\cite{LandoltBornstein2001_2}. As shown in Table \ref{res_bulkLat_R}, all DFT functionals except the ferromagnetic solutions for PBE+U and SCAN+U correctly predict the VO\tsub{2} phase to be metallic. 
In the latter two cases, a \SI{0.42}{eV} and a \SI{1.04}{eV} band gap is opened for structurally relaxed cell geometries. Quite clearly the opening of a gap also depends on structural details since no gap is opened using PBE+U at U~=~\SI{2}{eV} and experimental structural data\cite{Stahl2019}. Interestingly, the gap in semiconducting V\tsub{2}O\tsub{5} is well described by all used functionals, with the largest deviation, \SI{0.4}{eV},  found for SCAN+rVV.
\par 
In order to compare the performance of all functionals and spin configurations, two more key figures are displayed in Table \ref{res_bulkLat_R}:
 the magnetization energy $\Delta E_{\text{FM-NM}}$  for rutile VO\tsub{2} and the reaction enthalphy for V\tsub{2}O\tsub{5} with respect both to nonmagnetic and ferromagnetic rutile VO\tsub{2} ($H_f^{\mathrm{V_2O_5}}$). 
Table \ref{res_bulkLat_R} shows that the calculated reaction enthalpy is always lower for the non spin-polarized calculations due to constraining all magnetic moments in the VO\tsub{2} phase to zero, which costs \SIrange{79}{720}{meV} per formula unit depending on the chosen DFT functional. The PBE functional yields the smallest value for $\Delta E_{\text{FM-NM}}$, and is increased by \SI{136}{meV} for SCAN and by \SI{242}{meV} for PBE+U.  SCAN+U  combines both effects and therefore leads to the largest value of  \SI{720}{meV} for $\Delta E_{\text{FM-NM}}$. As a result,  non-magnetic 
calculations  overestimate the enthalpy of reaction for V\tsub{2}O\tsub{5}  by \SIrange{520}{790}{meV}. Employing spin-polarized PBE+U, SCAN and SCAN+rVV functionals reduces this error to  \SIrange{100}{180}{meV} per V\tsub{2}O\tsub{5} unit. The PBE functional overestimates $H_f^{\mathrm{V_2O_5}}$  even for a ferromagnetic treatment, which is the consequence of two facts: First, the heat of formation is the lowest among the used functionals when performing non-magnetic calculations and secondly, the magnetization energy for the PBE functional is significantly lower compared to  PBE+U and SCAN. On the other hand, the spin-polarized SCAN+U functional strongly reduces $H_f^{\mathrm{V_2O_5}}$ as compared to SCAN and 
underestimates the experimental value by \SI{890}{meV}. A similar decrease of $H_f^{\mathrm{V_2O_5}}$ is also visible inspecting the values for the PBE and PBE+U functionals, where the PBE value is quite strongly decreased by taking into account on-site V-d coulomb interactions.   

\begin{table*}
\setlength{\tabcolsep}{5pt}
\begin{tabular}{lc|c|c|c|c|c||c|c|c|c|c|c}
& \multicolumn{6}{c||}{VO\tsub{2} (rutile)} & \multicolumn{6}{c}{V\tsub{2}O\tsub{5}}   \\
 & $a$ & $c$ & $c/a$ & $V$ & $E_g$ & $\Delta E_{\text{FM-NM}}$ & $a$ & $b$ & $c$ & $V$ & $E_g$ & $H_f^{\mathrm{V_2O_5}}$ \\ 
\hline 
(NM) PBE & 4.62 & 2.78 & 0.60 & 29.6 & 0 &   & 11.54 & 3.57 & 4.88 & 50.2 
& 2.0 & -2.07 \\ 
(NM) PBE (ref. \cite{das_structural_2019}) &  \multicolumn{6}{c||}{--} & 11.55  & 3.57  & 4.72  & 48.6   & 1.98   & --   \\ 
(NM) PBE+U & 4.63 & 2.79 & 0.60 & 30.0 & 0 &   & 11.52 & 3.61 & 4.86 & 50.5 & 2.2 & -2.02 \\ 
(NM) SCAN & 4.56 & 2.77 & 0.61 & 28.8 & 0 & --  & 11.59 & 3.55 & 4.24 &  43.6 & 2.0 & -1.89 \\ 
(NM) SCAN (ref. \cite{Stahl2019}) & 4.57 & 2.77 & 0.61 & 28.9 & 0 &   & \multicolumn{5}{c|}{--}  &--   \\ 
(NM) SCAN+U & 4.59 & 2.76 & 0.60 & 29.1 & 0 &   & 11.59  & 3.58   & 4.20  
& 43.7  & 2.2  &  -1.83  \\ 
(NM) SCAN+rVV & 4.55 & 2.76 & 0.61 & 28.6 & 0 &   & 11.63 & 3.55 & 4.14 & 
42.7 & 1.9 & -1.80 \\ 
\multicolumn{12}{c}{} \\
\hline
\hline
\multicolumn{12}{c}{}
\\
(FM) PBE & 4.59 & 2.84 & 0.62 & 29.9 & 0 & -79 &   \multicolumn{5}{c|}{} & -1.91 \\ 
(FM) PBE+U & 4.50 & 3.02 & 0.67 & 30.6 & 0.42 & -321 &   \multicolumn{5}{c|}{} & -1.38 \\ 
(FM) SCAN & 4.54 & 2.83 & 0.62 & 29.1 & 0 & -215 & \multicolumn{5}{c|}{--} & -1.46 \\ 
(FM) SCAN (ref. \cite{Stahl2019}) & 4.54  & 2.83 & 0.62 & 29.1 & 0 & --   
& \multicolumn{5}{c|}{} & --   \\ 
(FM) SCAN+U & 4.46 & 3.00 & 0.67 & 29.8 & 1.04 & -720 &\multicolumn{5}{c|}{} & -0.39   \\ 
(FM) SCAN+rVV & 4.53 & 2.83 & 0.62 & 29.0 & 0 & -212 &\multicolumn{5}{c|}{} & -1.38 \\ 
\multicolumn{12}{c}{} \\
\hline
\hline
\multicolumn{12}{c}{}
\\
Expt.\cite{Rogers1993,Enjalbert1986,LandoltBornstein2001_2} & 4.55 & 2.86 & 0.63 & 29.6 & 0 & --   & 11.51 & 3.56 & 4.37 & 44.7 & 2.3 & -1.28 \\ 
\end{tabular} 
\caption{Structural and energetic details  of  the rutile VO\tsub{2} and V\tsub{2}O\tsub{5} phases, 
calculated with several DFT functionals for ferromagnetic (FM) and non-magnetic (NM) spin configurations at respectively optimized relaxed structures. $c/a$ stands for the ratio of the $c$ and $a$ lattice parameters and $V$ for the volume of the unit cell (in
\si{\AA\cubed}) . $E_\mathrm{g}$ denotes the band gap  (in eV) and $\Delta E_{\text{FM-NM}}$ 
the magnetization energy (in meV) defined as energy difference between the FM and NM spin configuration.
$H_f^{\mathrm{V_2O_5}}$ is the reaction enthalpy for forming the V\tsub{2}O\tsub{5}  phase with 
respect to rutile VO\tsub{2}. The volumes, $\Delta E_{\text{FM-NM}}$ and $H_f^{\mathrm{V_2O_5}}$ are given per formula unit.} 

\label{res_bulkLat_R}
\end{table*}

\begin{figure}
\includegraphics[width=\linewidth]{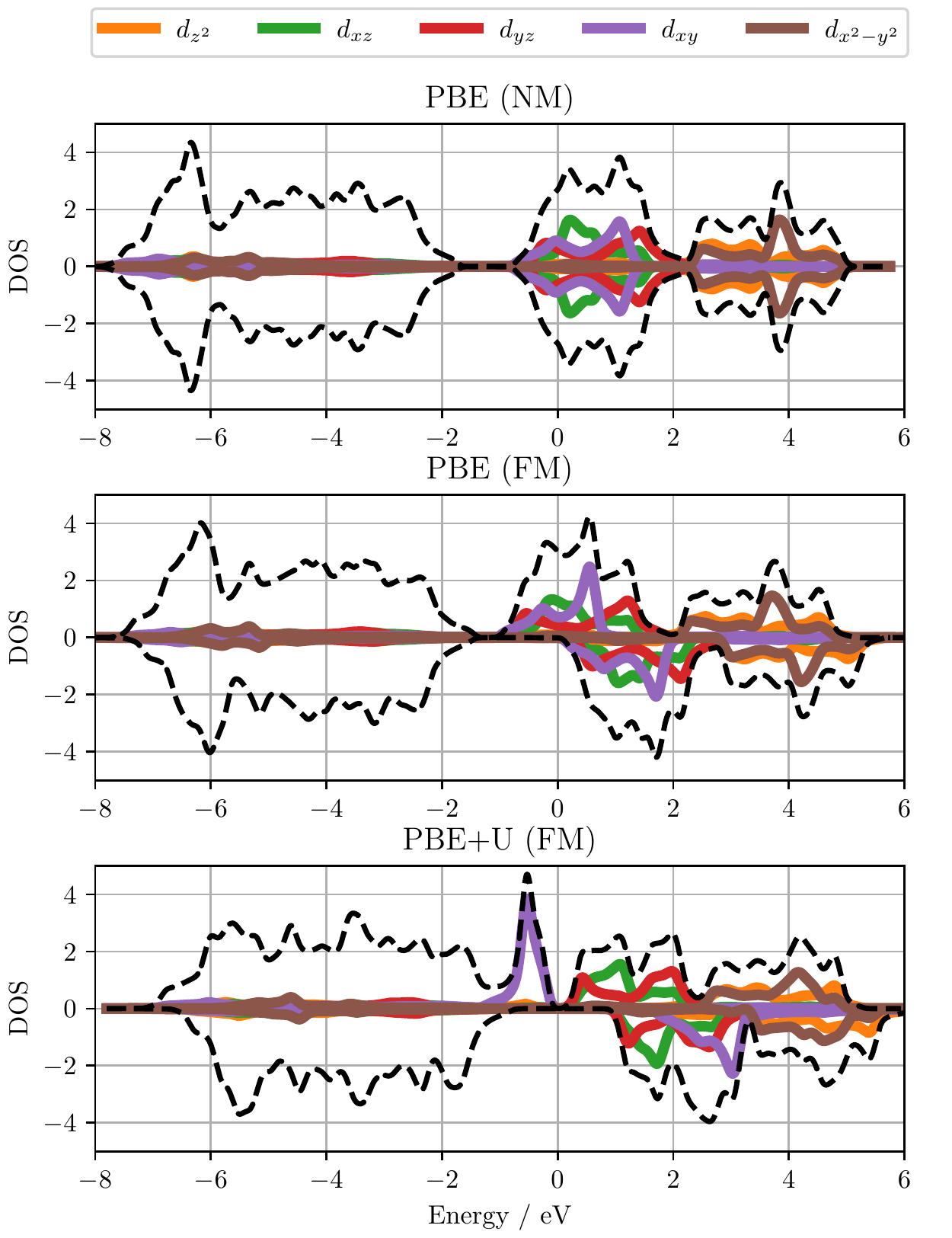}
\caption{Density of states (DOS) of the rutile (R) VO\tsub{2} phase calculated with different DFT functionals and projected onto Wannier orbitals localized on vanadium atoms.}
\label{fig:DOSR}
\end{figure}
\par
 Investigating the electronic structure of metallic rutile VO\tsub{2} with the help of Wannier
 projections onto atomic-like vanadium \textit{d} orbitals allows for a more
 explicit discussion. Figure \ref{fig:DOSR} shows the total density of states
 (DOS) and projections to vanadium atomic orbitals calculated with
 the above set of functionals and spin configurations. Since the 
 spin-polarized functionals, PBE, SCAN, SCAN+rVV yield similar
 projections as well as the PBE+U and SCAN+U we will only discuss
 the differences between the PBE and PBE+U functionals. Considering
 non-magnetic calculations, all functionals yield a  band structure similar to
 the one obtained from PBE calculations (see lowest panel in the
 Figure \ref{fig:DOSR}). As shown, the band structure includes oxygen
 $2p$ bands in a range from \SIrange{-8}{-1.6}{eV} and vanadium $3d$ bands 
 between  \SIrange{-1.6}{6}{eV}. Due to the
 crystal field the vanadium $d$ bands are split into lower $t_{2g}$ states (
 $d_{xz}$, $d_{yz}$ and $d_{xy}$ orbitals) and higher $e_g$ states (
 $d_{z^2}$ and $d_{x^2-y^2}$ orbitals). Including spin-polarization the
 splitting of the spin-up ($\uparrow$) and spin-down ($\downarrow$) channels,
 pushes the $t_{2g\downarrow}$ states above the Fermi level.
 Consequently, all electrons in the vanadium $3d$ bands are located in
 $t_{2g\uparrow}$ states. Including  on-site coulomb correlations via an Hubbard U splits  the $t_{2g\downarrow}$ further
and separates the $d_{xy}$ orbital
 from the $d_{xz}$ and $d_{yz}$ orbitals, and hence a \SI{0.42}{eV} energy gap is opened. 
 Here, all electrons in the V-$3d$ band are found in the $d_{xy}$ orbital which coincides with the edge of 
the O-$2p$ band. These results should be directly compared to state-of-the-art dynamical mean-field theory (DMFT) benchmark calculations 
as found e.g. in Fig.1 of Ref.\cite{Brito2016} (Note,  $d_{xy}$ in the present work corresponds to $a_{1g}$ in \cite{Brito2016}, $d_{xz}$  to  $e_g^{\pi}(1)$, $d_{yz}$ to $e_g^{\pi}(2)$, and the remaining orbitals to $e_g^{\sigma}$). 
A visual comparison directly reveals the improper description of the orbital occupancies for the spin-polarized PBE+U case, where a monoclinic-like $d_{xy}$ subband  is split off by \SI{0.42}{eV} from the $d_{xz}$ and $d_{yz}$ orbitals, rendering rutile VO\tsub{2} a semiconductor. This clearly contradicts the benchmark DMFT results for rutile VO\tsub{2} (R) where the DOS does not show this splitting and the system remains metallic (see upper panel (a) of Fig. 1 in \cite{Brito2016}).
However, this spurious behaviour is only observed for larger values of U (U~$\geq$~\SI{2}{eV}) and using the relaxed 
optimized lattice where the \textit{c}-direction comes out too large.
\par
The SCAN+U functional describes the structural and electronic properties of the rutile VO\tsub{2} even worse: The rutile \textit{c}-lattice
vector is severely overestimated which leads to an even stronger distorted unit
cell, the spurious band gap is increased to 
\SI{1.04}{eV}, the respective orbital occupations are in disagreement with DMFT benchmark calculations and the calculated reaction enthalpy $H_f^{\mathrm{V_2O_5}}$ deviates by more than \SI{200}{\percent} from experimental value (\SI{1.3}{eV}). Therefore, the SCAN+U functional has been discarded in the subsequent surface studies.
\subsection{VO\tsub{2} surface terminations}
\begin{figure*}
\centering
\includegraphics[width=\linewidth]{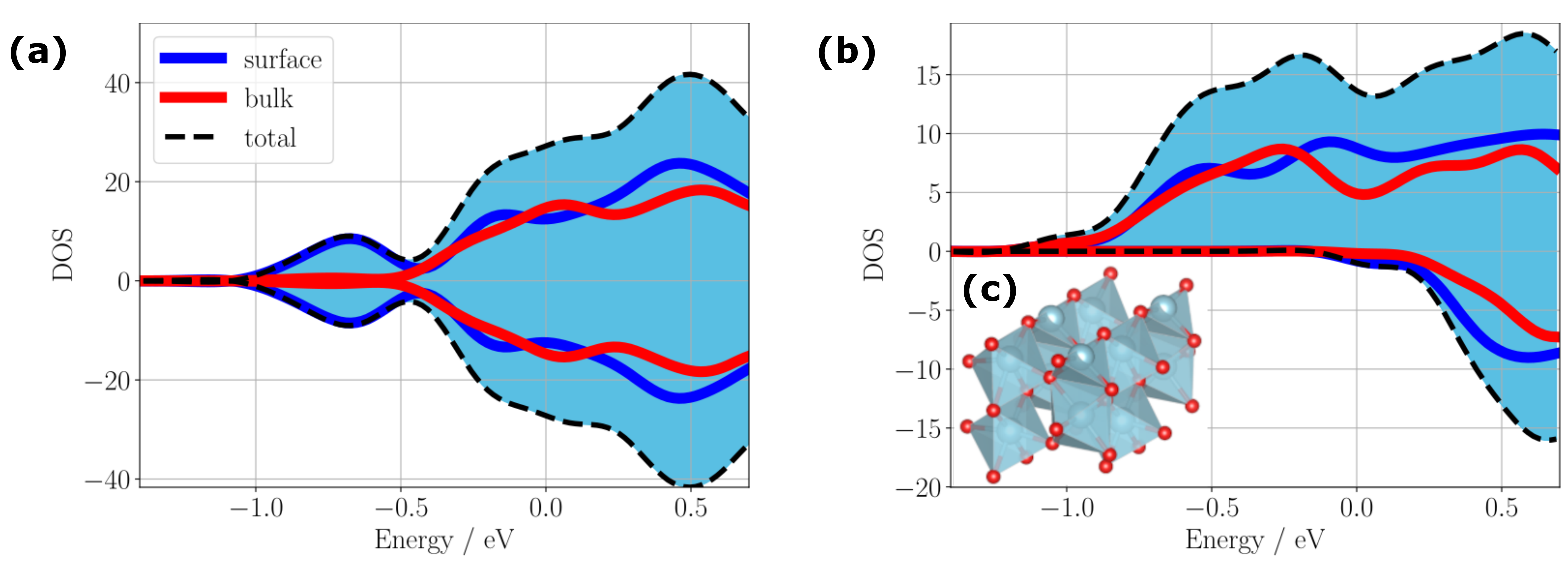}
\caption{Layer-resolved V-$t_{2g}$ DOS of a 7-layer VO\tsub{2} (001) slab and employing PBE. 
Surface denotes the DOS  projected onto V atoms located both in the surface and subsurface 
layers, while  bulk denotes the contributions of the three inner bulk-like layers. Panels (a) and (b) show the respective DOS for NM and FM spin configurations. The inset (c) shows the (001) surface structure comprising parallel chains of alternating 
VO\tsub{2} octahedra and tetrahedra.}
\label{fig:surfDOS}
\end{figure*}

In the following our evaluation of the performance of the various functionals together in conjunction with the treatment of the spin configuration is extended to surfaces. The primary focus is on the (110) surface but to allow for a comparison to previous work which used  (NM)
PBE\cite{Mellan2012} or (FM) PBE+U\cite{Wahila2020} functionals, also these terminations have been considered briefly. The calculated surface energies for five different surface terminations for our chosen DFT functionals and spin configurations are shown in Table
\ref{res_surfFreeEn}, including the already reported values for the PBE and PBE+U functionals. 
For the PBE functional, our values for the (100)  and (001) surfaces are in perfect agreement 
with Mellan and coworkers\cite{Mellan2012}, but the present work differs significantly, by up 
to \SI{25}{\percent}, for the more open (110) surface. The origin of this difference
can be traced 
 to a differing computational setup, namely a different number of core electrons in the chosen PAW potential, and  
 a different thickness of the slab. The surface free energies in \cite{Mellan2012} 
have been calculated using a four layer slabs only, whereas the slab thickness in our calculations was varied from five layers to eight layers to ensure a proper convergence of the surface energies.
For the (NM) PBE+U functional our results are again in a perfect agreement with the values reported by Wahila\cite{Wahila2020}, except for the (101) surface where our value is, surprisingly, larger by \SI{90}{\percent}. 
This difference is presumably due to a different choice of 
the $U$ value ($U=\SI{2}{eV}$ here and $U=\SI{3.25}{eV}$ in \cite{Wahila2020}),
which according to our experience opens a band gap in the middle of a (101) slab. 
Our results  imply that the calculated surface energies only depend slightly on the spin configuration, aside from two special cases. In the first case,
values calculated with the PBE+U functional depend much more strongly on the spin configuration, which we blame on the spurious electronic instability in the V-$t_{2g}$ bulk states for (FM) spin configurations, as discussed above. While the (NM) PBE+U functional yields 
an occupation of all the orbitals comprising the V-$t_{2g}$ bands, the (FM) PBE+U functional almost exclusively occupies only the V-$d_{xy}$ orbital, as shown in Figure \ref{fig:DOSR}. 
However, using unrelaxed slabs close to the respective bulk geometries 
surface energies are again much less affected by the spin configuration, resulting in only an \SI{13}{\percent} decrease for the (NM) configuration, which confirms that the surface free energy differences are indeed caused by the spurious electronic transition (gap opening) in the bulk and respectively optimized geometries.
In the second case,  (NM) non spin-polarized calculations show a systematic 
$\sim$\SI{20}{\percent} decrease of the (001) surface free energy for all the functionals. This decrease is not observed for unrelaxed slabs, where the calculated surface energies
are more or less the same for (NM) and (FM) spin configurations. 
The additional stabilization of the relaxed slabs
is accompanied by a change in the surface V-$t_{2g}$ states depicted in Figures
\ref{fig:surfDOS}(a-b). As shown in panel \ref{fig:surfDOS}(a), 
(NM) non spin-polarized calculations
show a peak centered 
 around \SI{-0.7}{eV} below $E_F$ formed almost exclusively by  V-$3d$ states located close to or at the surface. 
 This is not the case for (FM) spin-polarized calculations where the V-$3d$ DOS is quite similar for 
 both bulk and at the surface vanadium atoms as easily detectable in \ref{fig:surfDOS}(b)). 
As the (001) surface is the only surface termination which comprises parallel 
chains of alternating VO\tsub{2} 
octahedra and tetrahedra (see panel \ref{fig:surfDOS}(c)), 
we suggest that the change in the electronic structure is driven by this specific surface structure.
\par 
In summary, the values of the surface free energies shown  in Table
\ref{res_surfFreeEn} confirm that a VO\tsub{2}(110) surface termination is the most
stable one, independent of the chosen DFT functional and spin
configuration. PBE and (NM) PBE+U functionals yield rather low values, 
which is in line with a previous conclusion that  PBE  underestimates the 
surface energy for correlated materials\cite{DeWaele2016}. But all other tested functionals 
provide more reasonable values.
Concluding this paragraph, we note that a comparison of surface free energies unfortunately reveals a 
strong dependence on the chosen functional, which must be kept in mind especially when discussing absolute numbers. 
We also suppose that similar effects might be present  in other surface orientations not treated in the present work.
\begin{table}[!t]
\setlength{\tabcolsep}{5pt}
\begin{tabular}{cc|cc|cc|c|c}

&  & \multicolumn{2}{c|}{PBE} & \multicolumn{2}{c|}{PBE+U} & SCAN & SCAN+rVV \\ 
\hline 
\multirow{5}{*}{FM}  & (110) & 26 & & 44 & & 36 & 48\\ 
 					&  (101) & 51 & & 66 & & 71 & 82\\
 					&  (100) & 28 & & 51 & & 47 & 60\\ 	
 					&  (001) & 74 & & 76 & & 97 & 108\\  	
 					&  (111) & 67 & & 72 & & 89 & 99\\ 
\hline
\multirow{5}{*}{NM} &  (110) & 24 &18\tsup{a} & 25 &22\tsup{b} & 40 & 53\\ 
					&  (101) & 51 &46\tsup{a} & 56 &30\tsup{b} & 76 & 85\\
 					&  (100) & 27 &26\tsup{a} & 30 &29\tsup{b} & 50 & 64\\ 	
 					&  (001) & 59 &60\tsup{a} & 60 &59\tsup{b} & 76 & 88\\  	
 					&  (111) & 68 &78\tsup{a} & 64 & & 90 & 101\\  				\multicolumn{8}{l}{\tsup{a} ref. \cite{Mellan2012}}\\
\multicolumn{8}{l}{\tsup{b} ref. \cite{Wahila2020}}\\ 						
\end{tabular} 
\caption{Calculated surface free energies in \si{meV\per\AA\squared} for various VO\tsub{2} surface orientations.}
\label{res_surfFreeEn}
\end{table}
%
%
\subsection{VO\tsub{2}(110) ($2\times 2$) reconstructions}
\begin{figure}
\centering
\includegraphics[width=\linewidth]{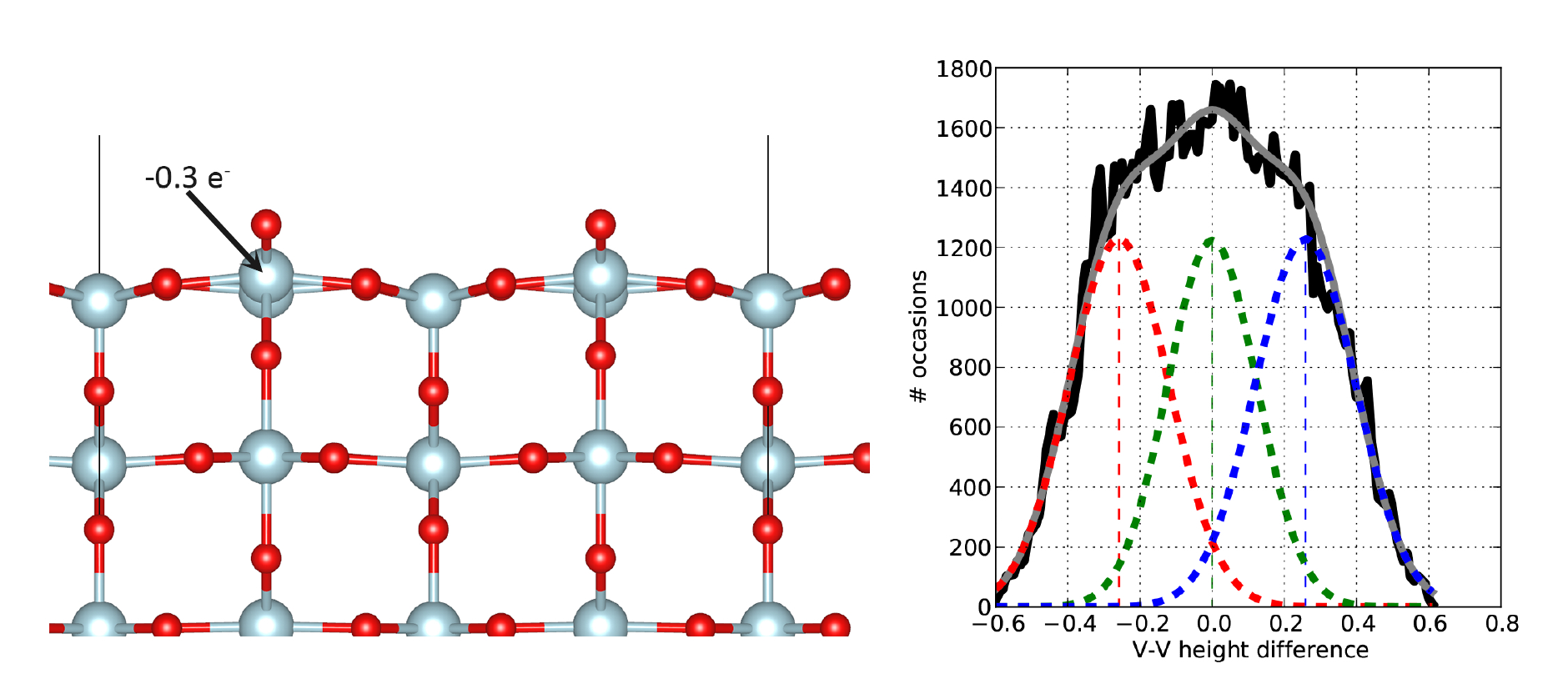}  
\caption{Left side: Buckled ($2\times 1$) reconstruction obtained from simulated 
annealing for non spin-polarized calculations. 
The size of the buckling in the surface layer amounts to  ~\SI{0.2}{\AA}, and is 
accompanied by a decrease of the occupation  (\SI{0.3}{electrons}) of the V\textit{3d} 
states of the upper vanadium atoms indicated by the black  arrow.
Right side: Distribution of height differences between the V-V pairs during 
a molecular dynamics simulation consisting of \SI{d5} steps with a time step of \SI{1}{fs} 
performed at \SI{80}{\celsius} (black line) including a fit with three Gaussian functions (red, blue and green lines) 
corresponding to the up-down, non-buckled and down-up configurations of the V-V pairs, respectively. The gray 
line is the sum of all fit functions.}
\label{fig:buckled}
\end{figure}

Interestingly, our results indicate that the calculated ground
state has a lower symmetry even for the bare (110) surface, 
resulting in the ($2\times 1$) buckled superstructure
shown in Figure \ref{fig:buckled}. The surface layer shows an 
unpaired zig-zag line pattern, similar to the monoclinic M2 phase\cite{Eyert2002}. 
The relative height difference in this
buckled row depends on the respective functional and ranges from
\SIrange{0.30}{0.43}{\AA}.
An instability towards buckling can already be observed for the bulk phases of
VO\tsub{2}: On the one hand, a V-V pairing is present along the [001]
direction of the monoclinic ground state. One the other hand, several DFT
functionals, such as PBE, predict an imaginary DFT phonon mode for the
rutile structure \cite{Kim2013}, where the oxygen atoms are shifted along 
the [001] direction with respect to the vanadium atoms. 
To exclude the possibility that the buckling at the surface is driven by this artificial bulk mode, 
we recalculated the structure where the bottom layer was
fixed at its bulk structure and the remaining atoms were not allowed to relax 
in the [001] and [1-10] direction.
Even under these constraints, the buckled surface has again a lower energy: 
buckling stabilizes the surface by \SIlist{40;15;9;11}{meV} per 
($2\times 1$) slab using the (NM) PBE, PBE+U, SCAN and SCAN+rVV functionals,
respectively. These results reinforce  the above assumption that this buckling is indeed a surface
effect and not driven by instabilities in the bulk phase. Fully relaxing the
slabs increases  the  stability induced by the buckling by
\SIlist{72;39;45;59}{meV} per the ($2\times 1$) slab, respectively, for the functionals above. 
The buckling
is accompanied by a change in the projected DOS onto V-\textit{3d}
orbitals of the upper vanadium atoms in the buckled row. The occupation
of the projected V-\textit{3d} DOS is decreased by $\sim$\SI{0.3}{electrons} (PBE)
as compared to an unbuckled surface. 
To the opposite, spin-polarized calculations with (FM) spin ordering reveal a different trend: 
while PBE, SCAN and SCAN+rVV
functionals strongly suppress the stability of the buckled surface and transform it back to a bare (110) surface, 
the buckling is recovered at the PBE+U level, with a \SI{26}{meV} energy gain for  fully relaxed slabs.
\par
Since the small energy gain due to the buckling might render an experimental
verification difficult, we performed a molecular dynamics (MD) simulation at experimental temperatures 
 to evaluate the thermal stability of the buckling. The simulation
for \SI{100}{ps} was performed employing the
non-magnetic PBE functional, which shows the largest energy difference between
the buckled and the bare (110) surface, a temperature of \SI{350}{K} and a time step of \SI{1}{fs}.
We monitored the most prominent feature of the buckled surface, namely possible switches in 
the z-positions (i.e. along the surface normal) of the V-V pairs.
In order to decrease the computational cost we reduced the thickness of the slabs to
four layers where the bottom layer was fixed and the k-points grid was reduced
to $2\times 2\times 1$ points. The results of the MD simulations are shown in
Figure \ref{fig:buckled} as the number of switches at the $y$ axis and the buckling size at the $x$ axis. The plot also
shows a fit of the MD data (black line) with three Gaussian functions, related
to the buckled (red line), non-buckled (green line) and flipped buckled (blue
line) configurations. The sum of the three Gaussian fits (gray line) shows
that the buckled configuration is thermally smeared since there are no maxima
at the positions of the Gaussian functions related to the buckled
configurations. Furthermore, the average flipping time from the up-down to
down-up configurations was calculated to be \SI{260}{fs}. Thus, we conclude that the
time dependent stability of the buckling is much shorter than the resolution limit 
of atomically resolved STM experiments, as imposed by the typical scanning rate.
\begin{figure*}
\centering
\includegraphics[width=\textwidth]{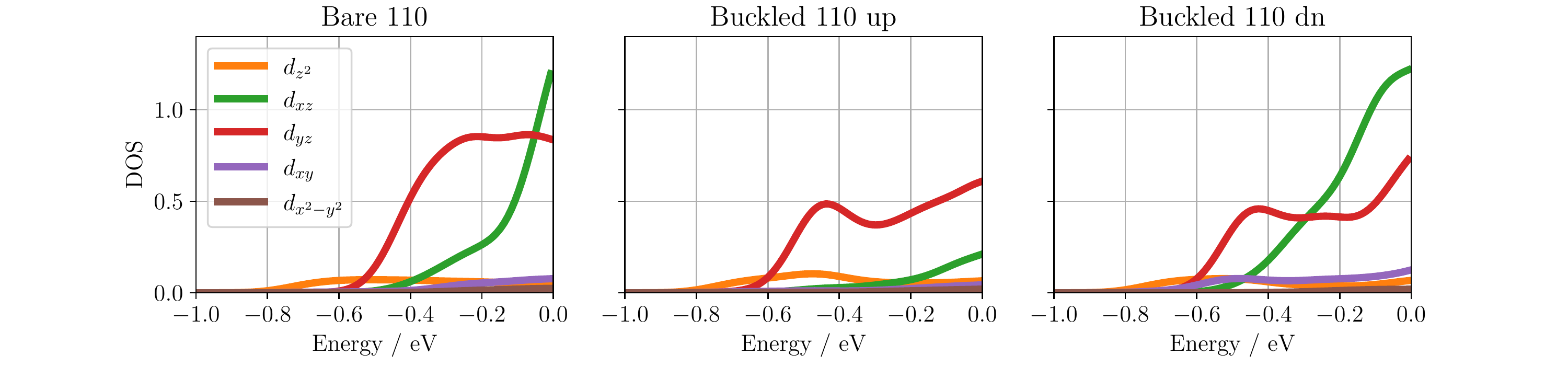}
\caption{(NM) non spin-polarized SCAN DOS projected onto Wannier functions 
localized at surface vanadium atoms of the bare  and  buckled (110) surface. Only the V-$3d$ contributions of the 
surface V atoms of the unbuckled (Bare(110)) and upper V (Buckled (110) up) respectively  lower V  (Buckled (110) dn) 
of the buckled VO\tsub{2} surface are shown. }
\label{fig:DOS_S_buckled}
\end{figure*} 
\begin{figure}
\centering
\includegraphics[width=\linewidth]{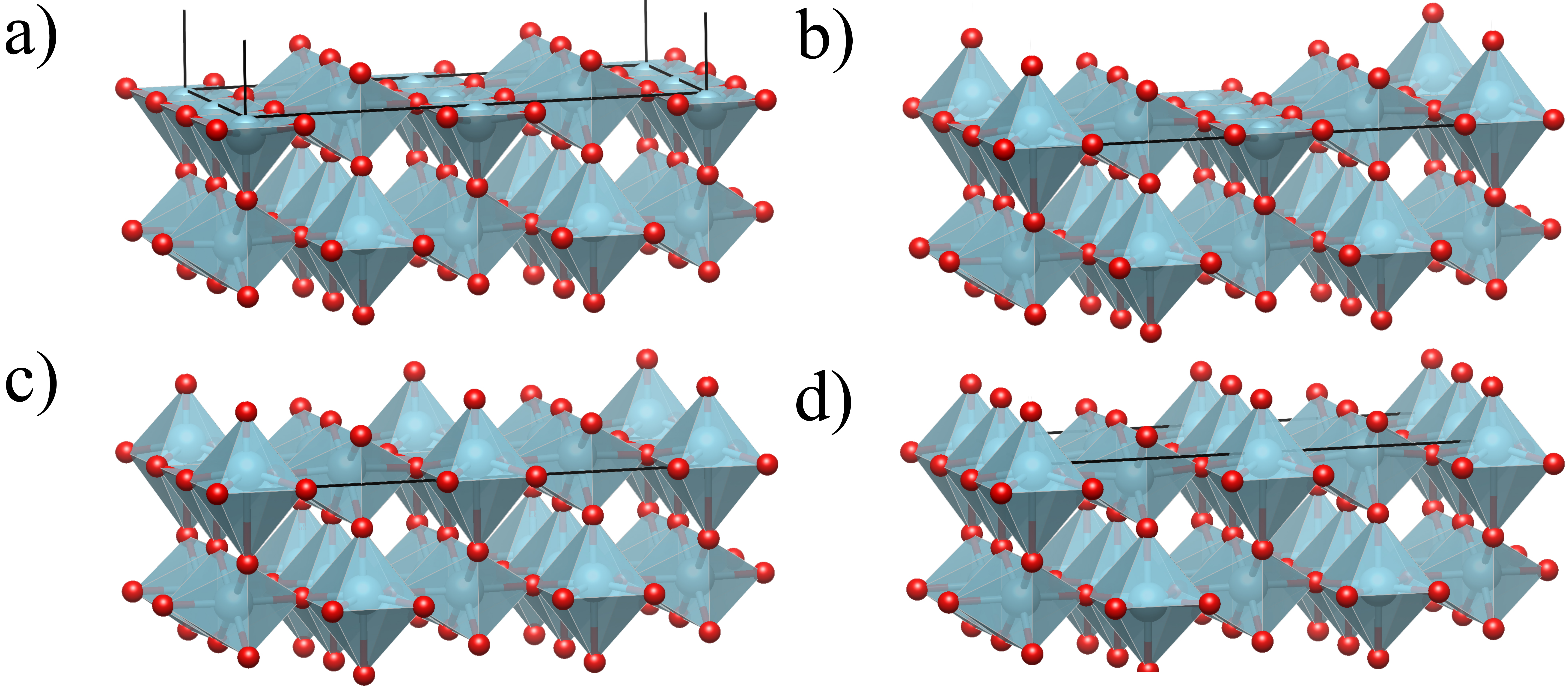}  
\caption{Perspective views of ($2\times 2$) supercells showing (a) the bare 110 surface, (b) 1 oxygen, (c) 2 oxygen (d) 4 oxygen adatoms per the ($2\times 2$) supercell.}
\label{fig:110StrucGeneral}
\end{figure}
\par
To shed some more light on the origins for the buckling we investigated the influence of the buckling on the
electronic structure with the help of Wannier projections.
Figure \ref{fig:DOS_S_buckled} shows the results using the SCAN functional. The left, middle and right panels show
projections onto  surface vanadium atoms at  unbuckled, buckled-up and
buckled-down sites. An inspection of these projections
reveals that a V-$d_{yz}$ orbital in the buckled surface is less occupied as compared to 
to the unbuckled surface and the occupation of a V-$d_{xz}$ orbital depends
on the z-position (up or dn) of the buckled vanadium atoms. 
The Wannier projections reveal that the impact of the surface buckling on the electronic structure is qualitatively different from changes induced by MIT. As it was pointed out by Eyert\cite{Eyert2002}, formation of vanadium pairs in the monoclinic phase is accompanied by an increase in the occupation number of the $d_{xy}$ orbital. Note that in the present work the notation of $d_{xy}$ and $d_{x^2-y^2}$ is interchanged. However, the buckled and the unbuckled surface show an opposite behavior: As shown in Figure \ref{fig:DOS_S_buckled}, the occupation number of the $d_{xy}$ orbital localized at both buckled and unbuckled sites is significantly reduced compared to the rutile bulk (see Figure \ref{fig:DOSR}, top panel).
\par
\begin{figure}
\centering
\includegraphics[width=\linewidth]{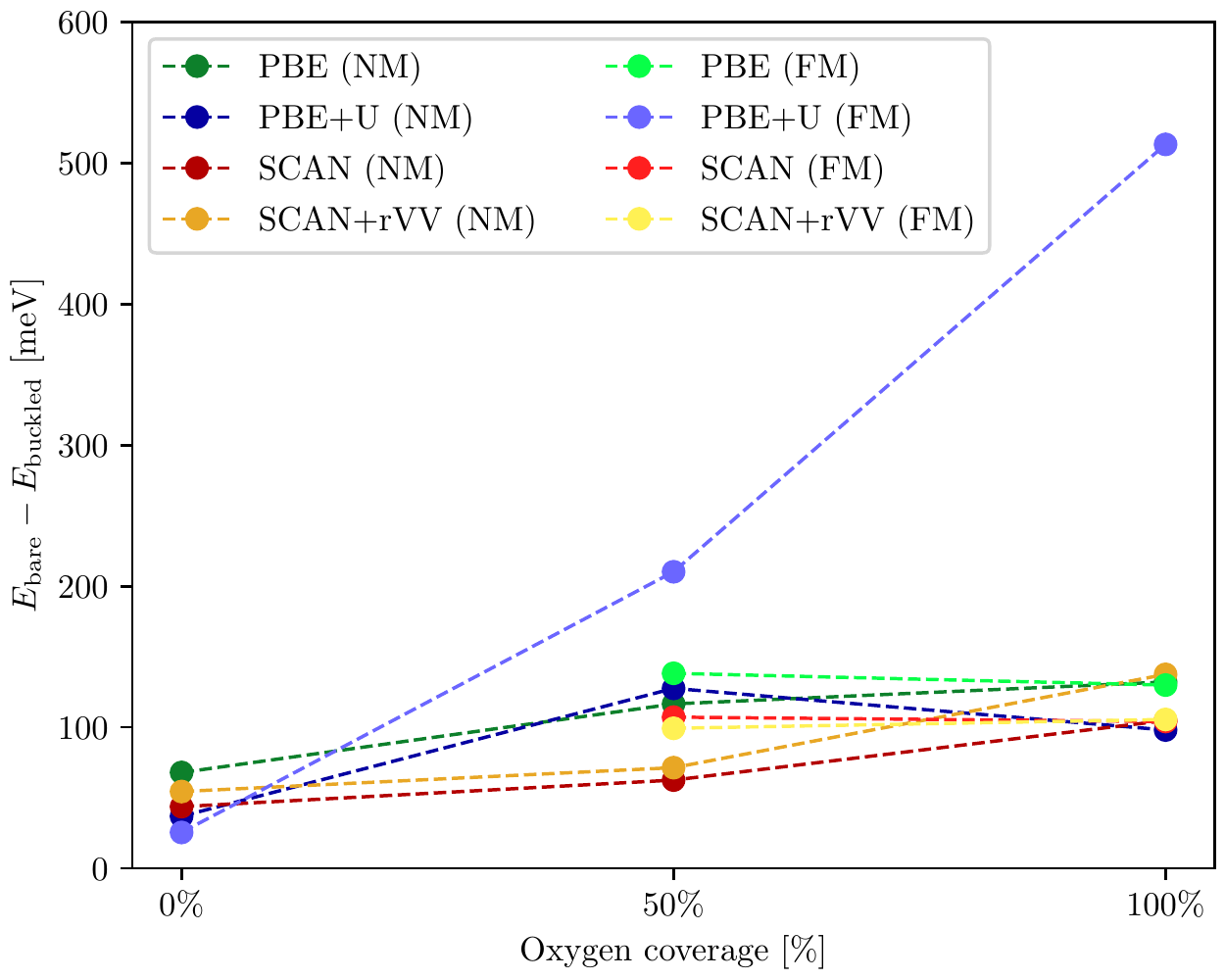}  
\caption{Additional stabilization (in meV per the ($2\times 1$) slab) induced by buckling in a VO\tsub{2}(110) surface, determined with several DFT functionals for non-magnetic (NM) and ferro-magnetic (FM) spin configurations. (FM) PBE, (FM) SCAN and (FM) SCAN+rVV predict no buckling for bare VO\tsub{2}(110) and therefore these points are not included.}
\label{fig:buckledStab}
\end{figure}

 
Since the occupation of the V-$3d$ states is quite sensitive to buckling, we also looked
into to what extent a varying oxygen coverage influences the size  of the buckling, comparing bare (clean), \SI{25}{\percent}, \SI{50}{\percent} and
\SI{100}{\percent} covered VO\tsub{2} surfaces as depicted in Figures
\ref{fig:110StrucGeneral}(a-d). Figure~\ref{fig:buckledStab} 
clearly shows that all non-magnetic functionals predict a
stable buckled surface 
and that its stability increases with oxygen coverage.
The energy gain due to the buckling ranges from \SI{39}{meV} (PBE+U) to \SI{72}{meV} (PBE) for the
bare surface, as already discussed above, and  increase with oxygen coverage, adding further
stability which ranges from \SI{61}{meV} (PBE+U, SCAN) to \SI{83}{meV}
(SCAN+rVV) per the ($2 \times 1$) slab  for a fully-covered surface. 
Quite evidently van der Waals interactions included in the
SCAN+rVV functional have only a negligible impact on the stability of the surface
buckling.

Since all spin-polarized functionals except  PBE+U predict an unbuckled surface, these negative values are omitted in Figure \ref{fig:buckledStab}, and only the (FM) PBE+U value of 
\SI{26}{meV} is included.
However, the trend changes with increasing the oxygen coverage to \SI{50}{\percent}, which compared to non-magnetic 
runs yields a further \SI{20}{\percent} (PBE), \SI{60}{\percent} (PBE+U), \SI{70}{\percent} (SCAN), and \SI{40}{\percent} (SCAN+rVV) stabilization of the surface buckling, which ranges from \SI{100}{meV} to \SI{140}{meV} per the ($2\times 1$) slab on an absolute scale. Increasing the oxygen coverage to the  fully-covered limit, the buckled  
surface is most preferred for the (FM) PBE+U functional which adds a \SI{510}{meV} extra stability per the ($2\times 1$) slab. The other functionals, PBE, SCAN and SCAN+rVV also prefer a buckled surface by \SI{104}{meV} to \SI{137}{meV} per the ($2\times 1$) slab without any significant difference between the non-magnetic and ferromagnetic spin configurations.
We attribute the comparably large energy gain for the (FM) PBE+U functional to the opening of the gap in the bulk and hence to a reduced screening of the induced charge imbalances upon oxygen adsorption. Hence, this buckling might play a non-negligible role for the stabilization of oxygen-rich surface reconstructions 
 proposed in recent experimental studies\cite{Fischer2020,Wahila2020} and even more on surfaces of the insulating  M phase of VO\tsub{2}.   
\par
\begin{figure*}
\centering
\includegraphics[width=\linewidth]{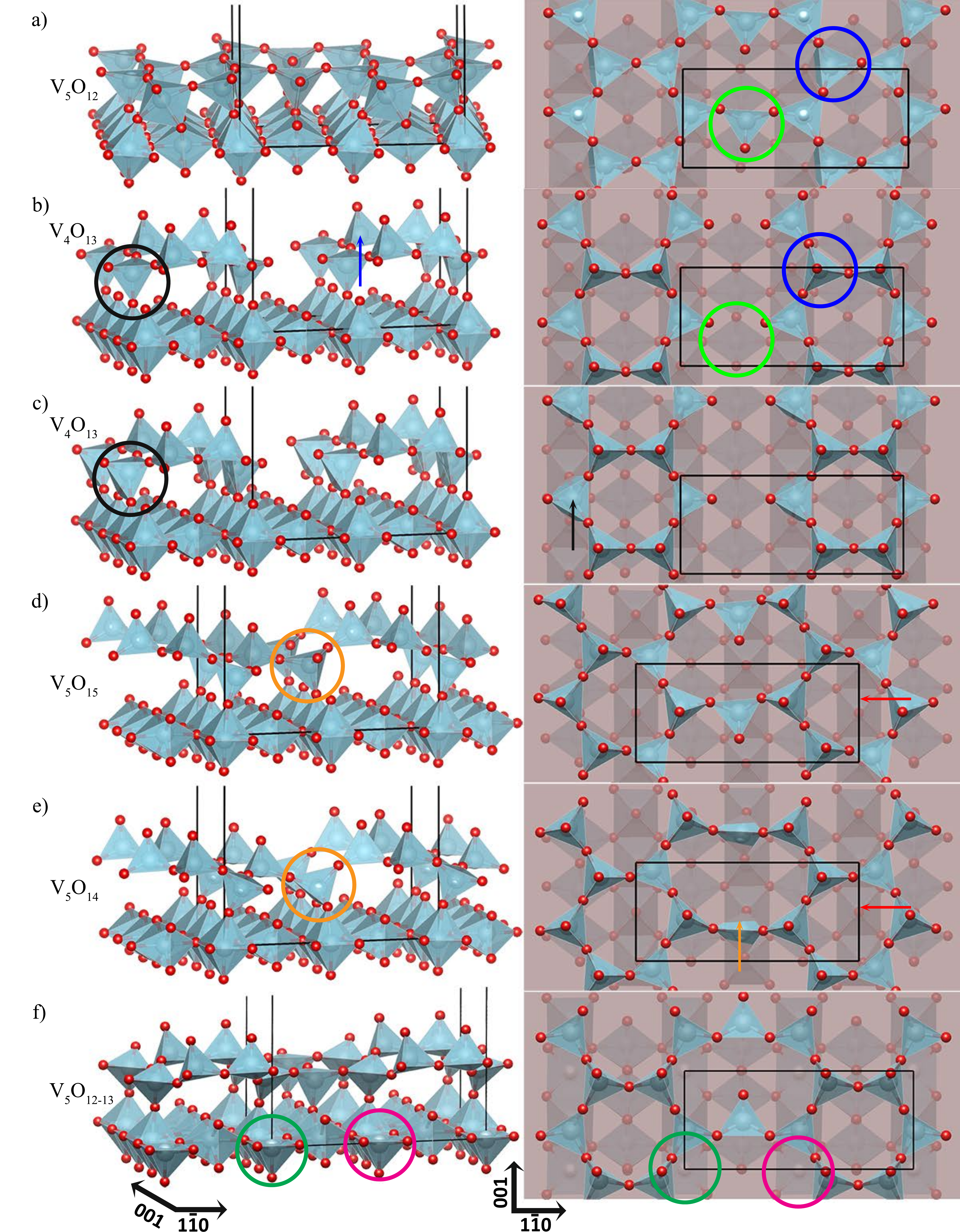}  
\caption{Perspective and top views of the resulting ($2\times 2$) superstructures obtained from an optimization of random structures with different surface stoichiometries. The structures descend from an underlying pattern shown in panel (a)  which is  structurally similar to the ($3\times 1$) reconstruction of the SrTiO\tsub{3}(110) surface \cite{Enterkin2010}.}
\label{fig_uspexStrucGeneral}
\end{figure*}      
\par

Since the experimentally observed ($2\times 2$) superstructure \cite{Wagner20}
cannot be explained  by the surface buckling only, we investigated the stability
of a large set of surface reconstructions obtained by optimizing preselected 
random structures for a number of surface stoichiometries and each employed 
functional and spin configuration. The detailed
procedure is outlined in the Supplementary Information. In the
following, a given surface stoichiometry is defined by the number of vanadium and oxygen 
atoms added on top of  bare rutile VO\tsub{2} (110) surface. 
 Independent of the chosen
functional and spin configuration,
the five most stable terminations which resulted 
from optimizing the preselected random structures 
are depicted in Figures \ref{fig_uspexStrucGeneral}(b-f). 
All of them are based on a general pattern shown in Figure
\ref{fig_uspexStrucGeneral}(a) which displays a 
V\tsub{5}O\tsub{12} surface stoichiometry and is structurally similar to the ($3\times 1$)
surface reconstruction of SrTiO\tsub{3}(110)\cite{Enterkin2010}. This
reconstructed surface superstructure consists of alternating octagonal and hexagonal rings made 
 of VO\tsub{2+x}
polyhedra as most easily visible in the top view of panel
\ref{fig_uspexStrucGeneral}(a). 
However, the  surface free energy that results from this termination is too high
so that a further oxidation is required  to stabilize
this superstructure by adding oxygen atoms in between the octahedral
rows of the subsurface layer (see the blue circle). The first stable
($2\times 2$) ring superstructure derived from the general pattern is shown in Figure \ref{fig_uspexStrucGeneral}(b).  
The modification proceeds 
in two steps: First, the bridging vanadium tetrahedra (see the green
circles) are removed and second, two additional oxygen atoms are adsorbed
on the ring pattern, which pushes the vanadium tetrahedra towards the vacuum
as indicated by the the blue arrow. These modifications 
lead to the formation of a stable, disconnected ring
structure with a V\tsub{4}O\tsub{13} surface stoichiometry and showing only VO\tsub{x} tetrahedra. 
This ring
termination is quite flexible and can be easily shifted by a quarter of the \textit{c}-bulk lattice vector 
as indicated by the black arrow in Figure 
\ref{fig_uspexStrucGeneral}(c). As a result, the lower VO\tsub{x} tetrahedra
are additionally bound to a second oxygen ad-atom which changes their
coordination geometries to square pyramids. This change is most easily visible 
by comparing the features within the black circles in Figures \ref{fig_uspexStrucGeneral}(b) and
(c). Both disconnected ring structures in Figures
\ref{fig_uspexStrucGeneral}(b,c) lead to a rectangular STM pattern (see Figure S2(a) in the Supplementary information)  where the bright features are due 
to the topmost oxygen atoms. The relative stability of these ring structures 
depends on the choice of the functional and will be discussed further below. 
Proceeding with the ring
structures, another variant is shown in Figure  \ref{fig_uspexStrucGeneral}(d),
which descends from the general pattern in the following way: First, the ring 
structure is shifted by a quarter of the [1$\bar{1}$0] bulk lattice vector (see the red arrow in the top view) so that the bridging tetrahedra are aligned with upright octahedra of the subsurface layer as marked by the orange circle. Second, this shift of the ring termination relative to the subsurface layer is accompanied by the oxidation of different vanadium tetrahedra.
All together this leads to a termination with an overall V\tsub{5}O\tsub{15} surface stoichiometry displaying a zig-zag STM pattern (see Figure S2(b) in the Supplementary information). 
The next stable superstructure shown in Figure \ref{fig_uspexStrucGeneral}(e) is generated by shifting the previous structure 
by an additional quarter of the [001] bulk lattice vector (see orange arrow), which again causes the
bridging tetrahedra to bind to two subsurface octahedra, as easily seen by comparing the features inside the orange 
circles in Figures \ref{fig_uspexStrucGeneral}(d,e). However, here, one oxygen atom is removed changing  the overall surface stoichiometry to V\tsub{5}O\tsub{14} and causing the tetrahedral coordination geometry not to switch to a square pyramid as has been the case for the V\tsub{4}O\tsub{13} ring structures. 
\par
Since the ring terminations are supported on
a 
bare surface,  also oxygen atoms might  be missing in the connecting subsurface layer
which changes the local coordination of the affected vanadium atoms from octahedra
to square pyramids. This is shown in Figure \ref{fig_uspexStrucGeneral}(f)
where we consider a ring structure similar to the shifted V\tsub{4}O\tsub{13}
ring (see Figure \ref{fig_uspexStrucGeneral}(c)) where the hexagonal rings are
connected by bridging tetrahedra. The
side view shows the stepwise removal of subsurface oxygen atoms surrounding two vanadium sites (see the green and violet circles in Figure
\ref{fig_uspexStrucGeneral}(f)), which leads to a V\tsub{5}O\tsub{13} and a
V\tsub{5}O\tsub{12} surface stoichiometry, respectively.   
\par
Our results reveal that the relative stability of the two V\tsub{4}O\tsub{13}
surface terminations shown in Figures \ref{fig_uspexStrucGeneral}(b,c) 
depends on the chosen functional. The energetic differences shown in Table \ref{res_ringComp} clearly indicate that the (FM) spin-polarized PBE and PBE+U functionals prefer the purely tetrahedral ring structure (b) by \SI{114} and \SI{385}{meV} per ($2\times 2$) supercell while all the other functionals find the shifted modification (c) as more stable by \SIrange{4}{324}{meV}.

\begin{table}
\setlength{\tabcolsep}{5pt}
\begin{tabular}{l|c|c|c|c}

V\tsub{4}O\tsub{13} &  PBE & PBE+U & SCAN & SCAN+rVV \\ 
\hline 
FM  & -114   & -385 & 97 & 137\\ 
\hline
NM & 4 & 108 & 230 & 324\\  					
\end{tabular} 
\caption{Total energy differences (meV per ($2\times 2$) supercell) of the disconnected ring terminations with
V\tsub{4}O\tsub{13} surface stoichiometry depicted in Figures \ref{fig_uspexStrucGeneral}(b,c). 
A negative number means that the purely tetrahedral ring structure shown in Figure \ref{fig_uspexStrucGeneral}(b) 
is preferred.}
\label{res_ringComp}
\end{table}
\begin{figure*}
\centering
\includegraphics[width=0.85\textwidth]{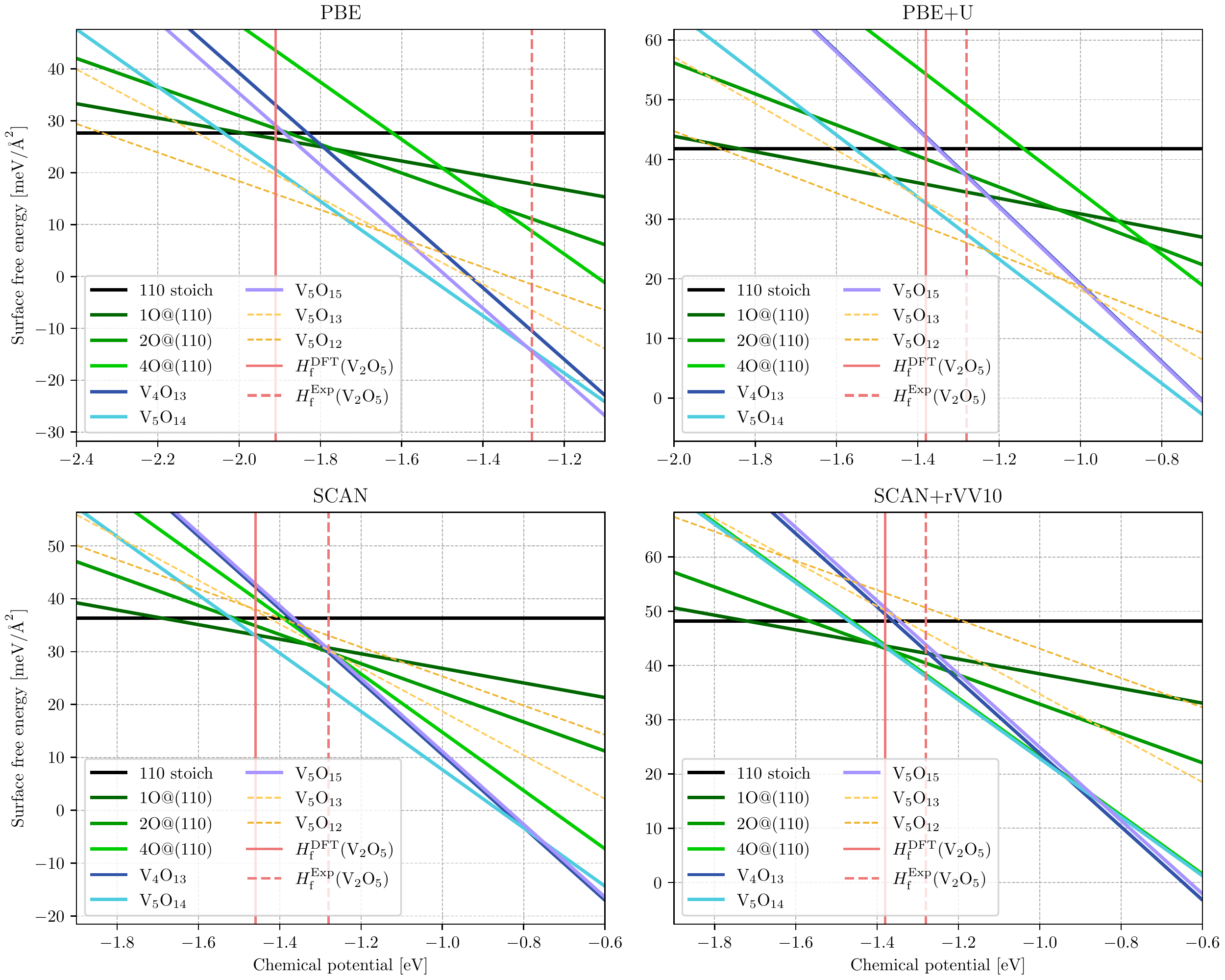}
\caption{Calculated surface free energies as a function of the oxygen chemical potential for various ($2\times 2$) VO\tsub{2}(110) surface terminations and a FM spin configuration. }
\label{fig:SurfFreeEn_mag}
\end{figure*} 
\par
\begin{figure*}
\centering
\includegraphics[width=0.85\textwidth]{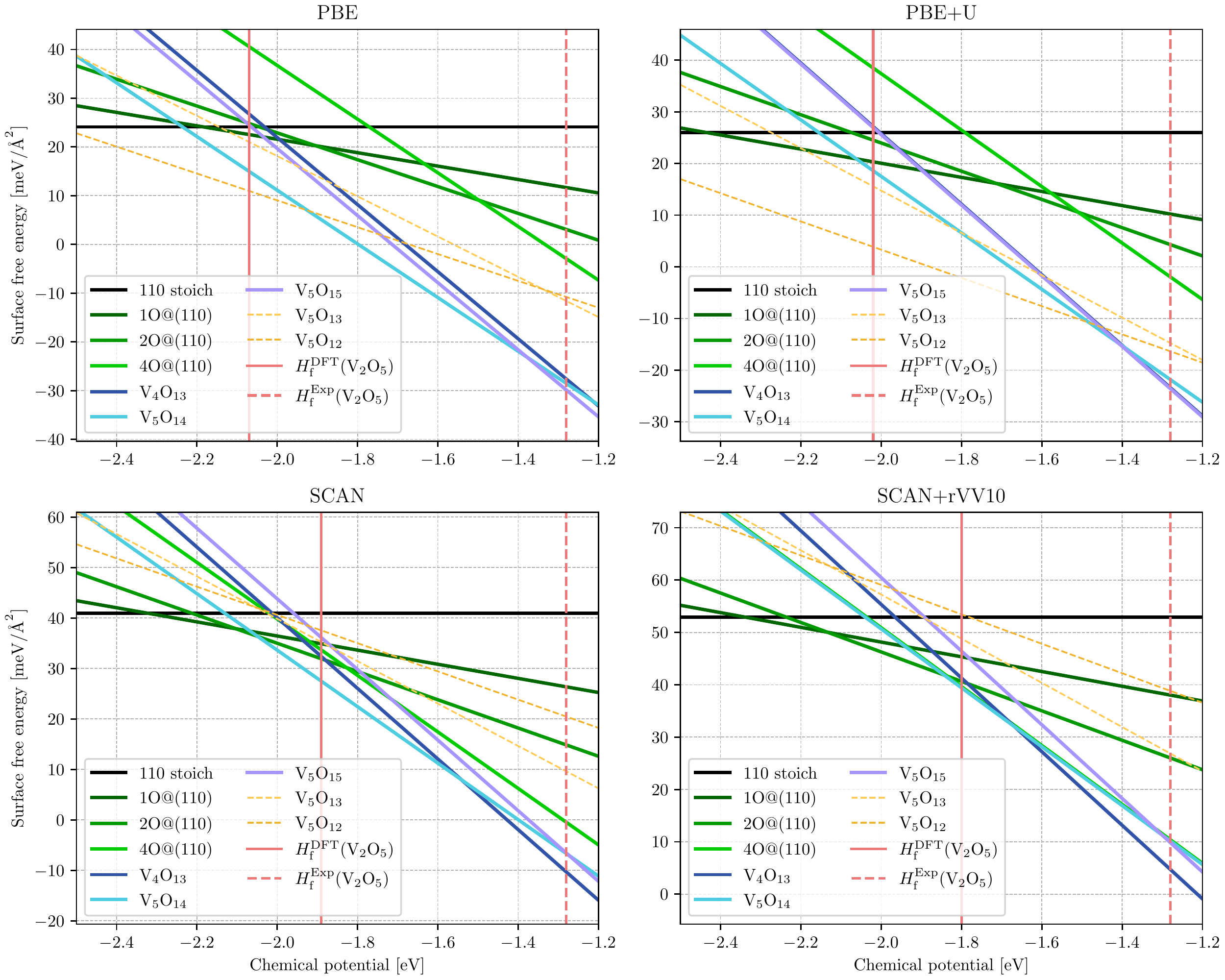}
\caption{Calculated surface free energies as a function of the oxygen chemical potential for various ($2\times 2$) VO\tsub{2}(110) surface terminations and a NM spin configuration.}
\label{fig:SurfFreeEn_nomag}
\end{figure*}

In order to compare the stability of the superstructures with different surface stoichiometries, we plot the surface free energy of the most stable terminations in Figures \ref{fig:SurfFreeEn_mag} and \ref{fig:SurfFreeEn_nomag} for ferromagnetic (FM)  and nonmagnetic (NM) calculations, respectively. Furthermore, we  compare the ring terminations with a  bare (110) surface (black line) and simple oxygen ad-atom (110) surface models shown in Figure \ref{fig:110StrucGeneral}. The green '1O', '2O', '4O' lines indicate the number of oxygen ad-atoms in the ($2\times 2$) supercell, namely one oxygen (Figure \ref{fig:110StrucGeneral}(b)), two oxygen (Figure \ref{fig:110StrucGeneral}(c)) and four oxygen atoms (Figure \ref{fig:110StrucGeneral}(d)), respectively. Orange vertical lines mark the calculated and experimental oxidation enthalpy to V\tsub{2}O\tsub{5} below which the bulk VO\tsub{2} is thermodynamically stable. 

\par
Oxygen adsorption on a rutile (110) surface has been already studied by Mellan
et al.\cite{Mellan2012} performing non spin-polarized PBE calculations. In their
work, the $\Gamma = +1$ and $\Gamma = +2$ surfaces are identical to our
\SI{50}{\percent} and fully covered VO\tsub{2} (110) terminations 
depicted in Figures \ref{fig:110StrucGeneral}(c,d).
However, our calculated adsorption energies for both coverages are  $\sim$\SI{0.46}{eV} larger
than the corresponding (i.e. the uncorrected) values for the $\Gamma = +1$ and $\Gamma = +2$ 
surfaces (see Figure 8 in ref.\cite{Mellan2012}).
This rather large discrepancy in adsorption energy is caused again mainly by a different computational setup. The use of  differing PAW potentials causes a \SI{0.26}{eV} difference and the different slab thickness increasing this value by another \SI{0.07}{eV}. Furthermore, the models in the present work are additionally stabilized by the surface buckling, which has been 
neglected in the work of Mellan
et al.\cite{Mellan2012}  which increases the difference by  $\sim$\SI{0.06}{eV} for both coverages. The remaining small discrepancy is attributed to other effects emerging from the different computational setup.     

\par
Figures \ref{fig:SurfFreeEn_mag} and \ref{fig:SurfFreeEn_nomag} show the
calculated surface free energies of the bare VO\tsub{2} (110) surface (black
lines), the oxygen ad-atom phases with different oxygen coverage (green lines)
and the ring terminations depicted in Figures
\ref{fig_uspexStrucGeneral}(b-f) for varying oxygen chemical potential. 
Concerning the V\tsub{4}O\tsub{13} ring structure only the surface free energy of the
most stable variant is included.  Independent of the chosen spin configuration all 
graphs show that all the functionals considered determine the ring
terminations to be more stable than  (PBE, PBE+U, SCAN) or at least similarly stable
(SCAN+rVV) as the oxygen ad-atom phases over the wide range of chemical
potentials. However, the relative stability of the ring terminations with
respect to the oxygen ad-atom phases is functional dependent. 
To quantify this dependency we
consider the difference between the crossing points of  both the fully covered oxygen ad-atom (110)
surface (light green line) and the V\tsub{5}O\tsub{14} ring phase (cyan line)
with the bare stoichiometric (110) surface, which determine the effective adsorption
energy of the additional oxygen atoms that are incorporated in the
superstructure. Independent of the spin treatment, the PBE and PBE+U
functionals determines this difference to fall between \SIrange{0.4}{0.5}{eV}, 
SCAN prefers the V\tsub{5}O\tsub{14} ring by \SI{0.1}{eV} and the
SCAN+rVV functional renders the difference in crossing points to
below \SI{0.01}{eV}. The graphs also show that the (FM) PBE, (NM) PBE and (NM)
PBE+U functionals prefer the ring structures over the whole range of the
oxygen chemical potentials while the other functionals and spin configurations
yield an energy window in the low oxygen chemical potential region where the
oxygen ad-atom phases would be preferred. On the other hand, the (FM) SCAN, (FM)
SCAN+rVV and (NM) SCAN+rVV functionals show a preference for the ring
terminations only in regions where the VO\tsub{2} bulk phase is calculated to
be thermodynamically unstable. 
The strong stabilization of the ring structures
for the PBE and PBE+U functionals also hints towards an existence of additional polyhedral 
surface terminations that are more stable than the ad-atom phases. 
Regarding the V\tsub{4}O\tsub{13} and V\tsub{5}O\tsub{15} rings, most of the functionals do not exhibit a significantly  
different stability values except for the (NM) SCAN, SCAN+rVV and (FM) PBE. While the V\tsub{5}O\tsub{15} ring
is  more stable by \SI{50}{meV} (SCAN) and \SI{80}{meV} (SCAN+rVV)  than the 
disconnected ring, the (FM) PBE functional prefers the connected V\tsub{5}O\tsub{15} ring by \SI{50}{meV}. 
\par
For the ring structure on the reduced oxygen missing subsurface layer (panel
\ref{fig_uspexStrucGeneral}(f) we find again a strong dependency on the chosen
functional. The graphs \ref{fig:SurfFreeEn_mag} and \ref{fig:SurfFreeEn_nomag}
find no additional stability at the SCAN or SCAN+rVV level, while these structures 
are preferred at the PBE and PBE+U level
 under strongly reducing conditions, see yellow dashed lines. Particularly
 the (NM) PBE+U functional  prefers this structure for oxygen chemical potentials 
 in the range of \SIrange{-1.5}{-2.8}{eV}. A enhanced 
stability is also found at the (NM) and (FM) PBE level, showing that a
doubly reduced subsurface layer would be stable for an oxygen chemical potential 
below \SI{-1.9}{eV} and \SI{-1.8}{eV} respectively. A stability window 
ranging from \SI{-1.23}{eV} to \SI{-1.89}{eV} is also found for 
(FM) PBE+U setups. The qualitative agreement between the PBE and PBE+U functionals 
can be understood by considering the existence of an electronic gap for states connected with the 
surface layer which will be discussed in the following section. 
The other functionals determine these reduced subsurface layer structures to be unstable with
respect to both the other other ring structures and ad-atom phases. Comparing the
reduced V\tsub{5}O\tsub{13} and V\tsub{5}O\tsub{12} ring terminations reveal yet another 
functional dependent property. Considering
their crossing points with the bare (110) surface line, the PBE and PBE+U
functionals show crossing to occur at a lower oxygen chemical potential 
for the more reduced ring termination with a V\tsub{5}O\tsub{12} surface 
stoichiometry. However, these crossing points are found either roughly at the same oxygen
chemical potential (SCAN) or they are even reversed (SCAN+rVV), revealing a
counter-intuitive opposite trend with respect to the ad-atom phases,
see
green lines in Figures \ref{fig:SurfFreeEn_mag} and
\ref{fig:SurfFreeEn_nomag}.
Vacancies at PBE and PBE+U level stabilize the surface which is not the case for the SCAN functional.
\par 
One of the most striking differences between the spin-polarized and non spin-polarized calculations 
for the surface free energies concerns the absolute values of the effective adsorption energies. 
This difference can be illustrated best for the ring termination with a V\tsub{5}O\tsub{14} 
surface stoichiometry (Figure \ref{fig_uspexStrucGeneral}(e). While the non-magnetic calculations yield effective adsorption energies, depending on the chosen functional, between \SI{-1.88}{eV} to \SI{-2.06}{eV}, (FM) spin-polarized calculations shift 
the resulting values by \SI{0.20}{eV}, \SI{0.60}{eV}, \SI{0.61}{eV} and \SI{0.57}{eV} to the right (less negative values) using PBE, PBE+U, SCAN or SCAN+rVV. Note that similar shifts ($\pm$\SI{0.1}{eV}) are also observed for the ad-atom phases (green lines in Figures \ref{fig:SurfFreeEn_mag} and \ref{fig:SurfFreeEn_nomag}). This energy shift can be attributed to higher energies for bulk rutile VO\tsub{2} when performing non-spin polarized calculations which are due to the neglect of local magnetic moments, see Table \ref{res_bulkLat_R}.
\par
\begin{figure}
\centering
\includegraphics[width=0.45\textwidth]{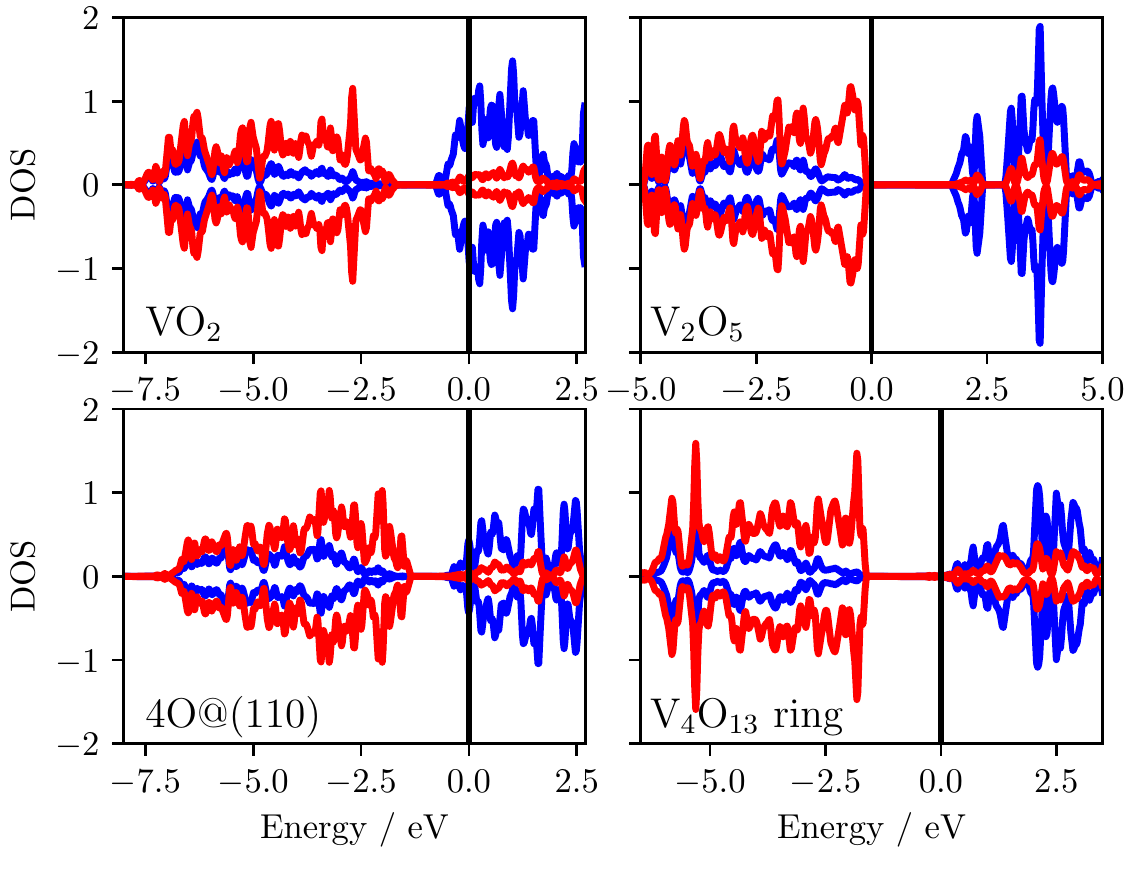}
\caption{Vanadium and oxygen projected DOS of the VO\tsub{2}, V\tsub{2}O\tsub{5} phases, and projections to surface 
atoms of a VO\tsub{2}(110) surface fully covered with oxygen ad-atoms and of the V\tsub{4}O\tsub{13} ring termination. The projected DOS was calculated with the non-magnetic PBE functional and is both normalized to a single atom and aligned at the upper edge of the O-$2p$ band.}
\label{fig:pDOS_ring}
\end{figure}

Considering a possible opening of a gap in the surface DOS of the oxygen rich ring terminations it is tempting to  inspect  the relation 
between the V\tsub{2}O\tsub{5} bulk phase and the
ring terminations more closely. 
In Figure \ref{fig:pDOS_ring} we compare the projected density of states of bulk VO\tsub{2} and
V\tsub{2}O\tsub{5} phases with the oxygen ad-atom phase at full coverage
(Figure \ref{fig:110StrucGeneral}(d)) and the V\tsub{4}O\tsub{13} ring. 
The major difference between the electronic structure of
VO\tsub{2} and V\tsub{2}O\tsub{5} bulk phases is found for the occupation of the V-$3d$ states. 
While the Fermi level cuts the V-$3d$ bands
in the VO\tsub{2} phase, the V\tsub{2}O\tsub{5} phase shows an
electronic band gap of about \SI{2.0}{eV} between the O-$2p$ and V-$3d$
states. As can been seen in Figure \ref{fig:pDOS_ring}, the projected surface DOS of the ring structure resembles  
rather closely the DOS of the V\tsub{2}O\tsub{5} bulk phase, revealing an unoccupied V-$3d$ band and hence this surface termination cannot form local magnetic moments. 
Figure \ref{fig:pDOS_ring} also shows that the terminating ring layer is insulating  with a \SI{1.9}{eV} band gap, which is in a good agreement with the \SI{2.3}{eV} obtained for the V\tsub{2}O\tsub{5} bulk  (see Table \ref{res_bulkLat_R}). 
It should also be noted that the surface free energies of the ring terminations
correlate with the calculated oxidation enthalpy of bulk VO\tsub{2}. As visible 
from the surface free energy graphs above, the difference between the effective
adsorption energy of the V\tsub{5}O\tsub{15} ring phase and the reaction
enthalpy for a V\tsub{2}O\tsub{5} formation by oxidizing  VO\tsub{2} is below
\SI{0.1}{eV} in all cases, while the absolute errors in calculated oxidation enthalpies range from
\SI{0.1}{eV} to \SI{0.79}{eV}. 
These facts indicate, that the ring terminations are related to a monolayer of V\tsub{2}O\tsub{5}(001) in agreement with our previous work \cite{Wagner20}. This finding also implies that a correct evaluation of the stability for both phases is important to properly describe the energetics of the oxygen rich surface terminations, a criterion 
which seems to be matched best by spin-polarized PBE+U, SCAN and SCAN+rVV calculations. However, the application of the  (FM) PBE+U  functional yields wrong orbital occupations and strongly overestimates the $c/a$ ratio of the bulk rutile VO\tsub{2} phase. From this perspective the spin-polarized SCAN and SCAN+rVV functionals seem to be a better choice for performing  calculations on VO\tsub{2} surfaces. 

%% file: summary.tex
\section{Summary}
In this work we studied the performance of  spin polarized and non-spin polarized GGA, GGA+U meta-GGA and meta-GGA+U functionals used to characterize bulk properties, stoichiometric surface terminations and oxygen-rich ($2 \times 2$) (110) reconstructions of the rutile VO\tsub{2} phase. We find that surface energies of particular 
surface orientations depend on the spin treatment due to complex electronic structure of both the surface states or the bulk states located in $t_{2g}$ orbitals. 
Considering a stochiometric  (110) surface, we 
identified a ground state with a buckled surface layer, resulting in a  ($2\times 1$) superstructure. The relative stability of the buckling depends on the chosen functional and becomes larger upon increasing the oxygen coverage which in turn reduces the occupation of the V\textit{3d} band near the surface. At high oxygen coverage, the (110) surfaces are terminated with V\tsub{2}O\tsub{5} related ring structures. 
We identified three important factors that influence the calculated stability of the these reconstructions. First,  the PBE and PBE+U functionals prefer quite open terminations loosely based on tetrahedral VO\tsub{2} polyhedra. This preference 
is already visible in the bulk calculations of the vanadium pentoxide phase, where the layer spacing is overestimated while the SCAN and SCAN+rVV functionals underestimate it (see Table \ref{res_bulkLat_R}). The leads to a large number of polyhedral superstructures that are stable at the PBE level, but unstable with the more advance SCAN meta-GGA functional. In addition, our results show 
that the V\tsub{5}O\tsub{15} ring termination is closely related to the vanadium pentoxide, which indicates that a correct description of the relative stabilities of the rutile VO\tsub{2} and V\tsub{2}O\tsub{5} phases plays an important role for the calculation of the  surface free energies. 
Our data show that  spin-polarized calculations are superior to the non-spin polarized ones for the oxygen covered surfaces, yielding in all cases (except the SCAN+U) the stability of the V\tsub{2}O\tsub{5} phase with respect to VO\tsub{2} 
closer to the experimental findings. The third factor that 
modifies the surface free energy 
is related to the surface buckling. Yet, while this buckling can change the adsorption energy by as much as 
0.5 eV  for a (2$\times$1) surface cell when using PBE+U Hubbard corrections for a (FM) spin configuration, 
this effect is much less pronounced for the other functionals. 
In summary, 
a ring structure reconstruction of the (110) surface is either energetically preferred or at least as stable as the adsorption phase at high oxygen chemical potential, independent of the chosen functional and spin treatment.

%% file: acknowledgement.tex
\ack
This work was supported by the Austrian Science Fund (FWF project F45), 
the Vienna Science and Technology Fund (WWTF), the City of Vienna and Berndorf Privatstiftung through project MA 16-005. We are grateful to the Vienna Scientific Cluster (VSC) for computational resources.

%% file: bibliography.tex
\bibliography{literature}

%% file: VO2_paper.bbl
\providecommand{\newblock}{}
\begin{thebibliography}{10}
\expandafter\ifx\csname url\endcsname\relax
  \def\url#1{{\tt #1}}\fi
\expandafter\ifx\csname urlprefix\endcsname\relax\def\urlprefix{URL }\fi
\providecommand{\eprint}[2][]{\url{#2}}

\bibitem{Stefanovich2000}
Stefanovich G, Pergament A and Stefanovich D 2000 {\em Journal of Physics:
  Condensed Matter\/} {\bf 12} 8837--8845 ISSN 0953-8984, 1361-648X

\bibitem{Strelcov2009}
Strelcov E, Lilach Y and Kolmakov A 2009 {\em Nano Letters\/} {\bf 9}
  2322--2326

\bibitem{Sengupta2011}
Sengupta S, Wang K, Liu K, Bhat A~K, Dhara S, Wu J and Deshmukh M~M 2011 {\em
  Applied Physics Letters\/} {\bf 99} 062114

\bibitem{Hiroi2013}
Hiroi Z, Hayamizu H, Yoshida T, Muraoka Y, Okamoto Y, Yamaura J~i and Ueda Y
  2013 {\em Chemistry of Materials\/} {\bf 25} 2202--2210

\bibitem{Netsianda2008}
Netsianda M, Ngoepe P~E, Catlow C~R~A and Woodley S~M 2008 {\em Chemistry of
  Materials\/} {\bf 20} 1764--1772 ISSN 0897-4756, 1520-5002

\bibitem{Zhu2012}
Zhu Z and Schwingenschl{\"o}gl U 2012 {\em Physical Review B\/} {\bf 86} 075149

\bibitem{Eyert2002}
Eyert V 2002 {\em Annalen der Physik\/} {\bf 11} 650--704

\bibitem{Wentzcovitch1994}
Wentzcovitch R~M, Schulz W~W and Allen P~B 1994 {\em Physical Review Letters\/}
  {\bf 72} 3389--3392

\bibitem{Liu2010}
Liu G~H, Deng X~Y and Wen R 2010 {\em Journal of Materials Science\/} {\bf 45}
  3270--3275

\bibitem{Stahl2019}
Stahl B and Bredow T 2020 {\em Journal of Computational Chemistry\/} {\bf 41}
  258--265

\bibitem{Bredow2018}
Bredow T and Stahl B 2018 {\em Chemical Physics Letters\/} {\bf 695} 28--33

\bibitem{Pantelides2017}
Xu S, Shen X, Hallman K~A, Haglund R~F and Pantelides S~T 2017 {\em Phys. Rev.
  B\/} {\bf 95}(12) 125105

\bibitem{Brito2016}
Brito W, Aguiar M, Haule K and Kotliar G 2016 {\em Physical Review Letters\/}
  {\bf 117} 056402

\bibitem{DMFTReview2006}
Kotliar G, Savrasov S~Y, Haule K, Oudovenko V~S, Parcollet O and Marianetti C~A
  2006 {\em Rev. Mod. Phys.\/} {\bf 78} 865--951

\bibitem{Mellan2012}
Mellan T~A and Grau-Crespo R 2012 {\em The Journal of Chemical Physics\/} {\bf
  137} 154706

\bibitem{Wahila2020}
Wahila M~J, Quackenbush N~F, Sadowski J~T, Krisponeit J~O, Flege J~I, Tran R,
  Ong S~P, Schlueter C, Lee T~L, Holtz M~E, Muller D~A, Paik H, Schlom D~G, Lee
  W~C and Piper L~F~J 2020 The {Breakdown} of {Mott} {Physics} at {VO}\tsub{2}
  {Surfaces} arXiv: 2012.05306 \urlprefix\url{http://arxiv.org/abs/2012.05306}

\bibitem{stahl_surfaces_2021}
Stahl B and Bredow T 2021 {\em ChemPhysChem\/} {\bf 22} 1018--1026

\bibitem{Wagner20}
Wagner M, Planer J, Heller B~S~J, Langer J, Limbeck A, Boatner L~A, Steinrück
  H~P, Redinger J, Maier F, Mittendorfer F, Schmid M and Diebold U 2021 {\em
  Physical Review Materials\/} {\bf submitted} (\textit{Preprint}
  \eprint{2107.00350}) \urlprefix\url{https://arxiv.org/abs/2107.00350}

\bibitem{Fischer2020}
Fischer S, Krisponeit J~O, Foerster M, Aballe L, Falta J and Flege J~I 2020
  {\em Crystal Growth \& Design\/} {\bf 20} 2734--2741

\bibitem{Kresse1996}
Kresse G and Furthm{\"u}ller J 1996 {\em Computational Materials Science\/}
  {\bf 6} 15--50

\bibitem{Zhang1996}
Zhang G~M and Hewson A~C 1996 {\em Physical Review B\/} {\bf 54} 1169--1186

\bibitem{Blochl1994}
Bl{\"o}chl P~E 1994 {\em Physical Review B\/} {\bf 50} 17953--17979

\bibitem{Joubert1999}
Kresse G and Joubert D 1999 {\em Physical Review B\/} {\bf 59} 1758--1775

\bibitem{Pack1977}
Pack J~D and Monkhorst H~J 1977 {\em Physical Review B\/} {\bf 16} 1748--1749

\bibitem{Pizzi2020}
Pizzi G, Vitale V, Arita R, Bl{\"u}gel S, Freimuth F, G{\'e}ranton G, Gibertini
  M, Gresch D, Johnson C, Koretsune T, Iba{\~n}ez-Azpiroz J, Lee H, Lihm J~M,
  Marchand D, Marrazzo A, Mokrousov Y, Mustafa J~I, Nohara Y, Nomura Y,
  Paulatto L, Ponc{\'e} S, Ponweiser T, Qiao J, Th{\"o}le F, Tsirkin S~S,
  Wierzbowska M, Marzari N, Vanderbilt D, Souza I, Mostofi A~A and Yates J~R
  2020 {\em Journal of Physics: Condensed Matter\/} {\bf 32} 165902

\bibitem{Sun2015}
Sun J, Ruzsinszky A and Perdew J 2015 {\em Physical Review Letters\/} {\bf 115}
  036402

\bibitem{Peng2016}
Peng H, Yang Z~H, Perdew J~P and Sun J 2016 {\em Physical Review X\/} {\bf 6}
  041005

\bibitem{Dudarev1998}
Dudarev S~L, Botton G~A, Savrasov S~Y, Humphreys C~J and Sutton A~P 1998 {\em
  Physical Review B\/} {\bf 57} 1505--1509

\bibitem{Wickramaratne2019}
Wickramaratne D, Bernstein N and Mazin I~I 2019 {\em Physical Review B\/} {\bf
  99} 214103

\bibitem{Reuter2001}
Reuter K and Scheffler M 2001 {\em Physical Review B\/} {\bf 65} 035406

\bibitem{LandoltBornstein2001}
{Collaboration: Scientific Group Thermodata Europe (SGTE)} 2001 Thermodynamic
  {Properties} of {Compounds}, {SbO\tsub{2}} to {Rh\tsub{2}O\tsub{3}} {\em Pure
  {Substances}. {Part} 4 \_ {Compounds} from {HgH}\_g to {ZnTe}\_g\/} vol 19A4
  ed {Lehrstuhl f{\"u}r Theoretische H{\"u}ttenkunde} and
  {Rheinisch-Westf{\"a}lische Technische Hochschule Aachen} (Berlin/Heidelberg:
  Springer-Verlag) ISBN 9783540410256

\bibitem{LandoltBornstein2001_2}
{Collaboration: Scientific Group Thermodata Europe (SGTE)} 2001 Thermodynamic
  {Properties} of {Compounds}, {V\tsub{2}O\tsub{5}} to {ThP} {\em Pure
  {Substances}. {Part} 4 \_ {Compounds} from {HgH}\_g to {ZnTe}\_g\/} vol 19A4
  ed {Lehrstuhl f{\"u}r Theoretische H{\"u}ttenkunde} and
  {Rheinisch-Westf{\"a}lische Technische Hochschule Aachen} (Berlin/Heidelberg:
  Springer-Verlag) ISBN 9783540410256

\bibitem{Oganov2006}
Oganov A~R and Glass C~W 2006 {\em The Journal of Chemical Physics\/} {\bf 124}
  244704

\bibitem{Lyakhov2013}
Lyakhov A~O, Oganov A~R, Stokes H~T and Zhu Q 2013 {\em Computer Physics
  Communications\/} {\bf 184} 1172--1182

\bibitem{Oganov2011}
Oganov A~R, Lyakhov A~O and Valle M 2011 {\em Accounts of Chemical Research\/}
  {\bf 44} 227--237

\bibitem{Rogers1993}
Rogers K~D 1993 {\em Powder Diffraction\/} {\bf 8} 240--244

\bibitem{Enjalbert1986}
Enjalbert R and Galy J 1986 {\em Acta Crystallographica Section C Crystal
  Structure Communications\/} {\bf 42} 1467--1469 ISSN 01082701

\bibitem{Goclon2009}
Goclon J, Grybos R, Witko M and Hafner J 2009 {\em Physical Review B\/} {\bf
  79} 075439 ISSN 1098-0121, 1550-235X

\bibitem{Kresse2001}
Kresse G, Surnev S, Ramsey M and Netzer F 2001 {\em Surface Science\/} {\bf
  492} 329--344 ISSN 0039-6028

\bibitem{das_structural_2019}
Das T, Tosoni S and Pacchioni G 2019 {\em Computational Materials Science\/}
  {\bf 163} 230--240

\bibitem{DeWaele2016}
De~Waele S, Lejaeghere K, Sluydts M and Cottenier S 2016 {\em Physical Review
  B\/} {\bf 94} 235418

\bibitem{Kim2013}
Kim S, Kim K, Kang C~J and Min B~I 2013 {\em Physical Review B\/} {\bf 87}
  195106

\bibitem{Enterkin2010}
Enterkin J~A, Subramanian A~K, Russell B~C, Castell M~R, Poeppelmeier K~R and
  Marks L~D 2010 {\em Nature Materials\/} {\bf 9} 245--248

\end{thebibliography}


\providecommand{\newblock}{}
\begin{thebibliography}{1}
\expandafter\ifx\csname url\endcsname\relax
  \def\url#1{{\tt #1}}\fi
\expandafter\ifx\csname urlprefix\endcsname\relax\def\urlprefix{URL }\fi
\providecommand{\eprint}[2][]{\url{#2}}

\bibitem{Oganov2006}
Oganov A~R and Glass C~W 2006 {\em The Journal of Chemical Physics\/} {\bf 124}
  244704

\bibitem{Lyakhov2013}
Lyakhov A~O, Oganov A~R, Stokes H~T and Zhu Q 2013 {\em Computer Physics
  Communications\/} {\bf 184} 1172--1182

\bibitem{Oganov2011}
Oganov A~R, Lyakhov A~O and Valle M 2011 {\em Accounts of Chemical Research\/}
  {\bf 44} 227--237

\end{thebibliography}
